\documentclass[twocolumn,twocolappendix,dvipsnames]{aastex702}
\usepackage{natbib}

\usepackage{graphicx}
\usepackage[utf8]{inputenc}

\usepackage{mathrsfs, amsmath, amssymb, amsthm, amsxtra, bm}
\usepackage{hyperref}
\usepackage[section]{placeins}

\usepackage{colortbl}
\usepackage{xcolor}

\usepackage{markcmds}
\usepackage{seancmds}
\usepackage{float}

\newcommand{\sorthelp}[1]{}

\newcommand{\red}[1]{#1}

\def\printmathfont#1#2{%
  \ifcat\relax#2\relax
    \count0=#2%
  \else
    \count0=\mathcode`#2%
  \fi
  \divide\count0 256%
  \count1=\numexpr\count0/16\relax                
  \count0=\numexpr\count0 - 16*(\count0/16)\relax 
  \def\tempinfo##1{\xdef\temp{\fontname#1##1}}%
  $%
  \ifnum\count1=7\relax
    \ifnum\fam>0\relax
      \ifnum\fam<16\relax
        \tempinfo{\fam}%
      \else
        \tempinfo{\count0}%
      \fi
    \else
      \tempinfo{\count0}%
    \fi
  \else
    \tempinfo{\count0}%
  \fi
  $%
  \string#1 \string#2 ($#2$): \temp
}

\begin{document}

\title{Simulating and Characterizing Non-Gaussian Small-Scale Polarized Emission}

\author[orcid=0009-0002-8645-5139]{Mark A. Bishop}
\email[show]{mark.bishop@vuw.ac.nz}
\affiliation{School of Chemical and Physical Sciences, 
Victoria University of Wellington,
Wellington 6012,
New Zealand}
\affiliation{Jet Propulsion Laboratory,
California Institute of Technology,
4800 Oak Grove Drive,
Pasadena,
CA 91109,
USA}

\author[orcid=0000-0001-7449-4638]{Brandon S. Hensley}
\email[show]{Brandon.S.Hensley@jpl.nasa.gov}
\affiliation{Jet Propulsion Laboratory,
California Institute of Technology,
4800 Oak Grove Drive,
Pasadena,
CA 91109,
USA}

\author[orcid=0000-0003-2933-8630]{Krzysztof M. Gorski}
\email{k.m.s.gorski@gmail.com}
\affiliation{Nicolaus Copernicus Acacdemy,
ul. Nowogrodzka 47a,
00-545 Warsaw,
Poland}

\author[orcid=0000-0002-5983-6481]{Charles Lawrence}
\email{Charles.Lawrence@jpl.nasa.gov}
\affiliation{Jet Propulsion Laboratory,
California Institute of Technology,
4800 Oak Grove Drive,
Pasadena,
CA 91109,
USA}

\author[orcid=0000-0002-4150-8076]{Graça Rocha}
\email{gra1roc2@gmail.com}
\affiliation{Nicolaus Copernicus Acacdemy,
ul. Nowogrodzka 47a,
00-545 Warsaw,
Poland}



\begin{abstract}






Galactic foregrounds present a major challenge for cosmic microwave background (CMB) cosmology as their amplitudes exceed the target primordial $B$-mode signal. While small-scale interstellar medium (ISM) emission is expected to be highly complex, current foreground simulations cannot reproduce the filamentary, non-Gaussian features of ISM structure in a stochastic way. To address this gap, we introduce a framework for generating stochastic small-scale foreground emission, using polarized synchrotron observations as an example use case. This method is capable of producing many independent realizations that additionally inherit morphological information from large angular scales. The underlying structural complexity incorporates anisotropic, multifractal, and advected non-Gaussian features that are tunable with a small number of parameters that carry physical interpretations. Because standard power spectra do not describe the complex phase coherence required for the structural complexity, we evaluate the synthetic fields using higher-order statistics, demonstrating that the Minkowski functionals and structure functions partially capture varying degrees of non-Gaussianity. This work lays the necessary foundation for new foreground simulation suites with stochastic realizations of possible non-Gaussian structure.

\end{abstract}

\section{Introduction}

Detection of the inflationary $B$-mode signal remains a primary goal for confirming the inflationary model and constraining the tensor properties of the Universe \citep{Kamionkowski.etal97, Hu.White97, Kamionkowski.Kovetz16}. However, this cosmological signal is heavily obscured by Galactic foregrounds. Synchrotron emission dominates the $B$-mode signal at low (radio) frequencies, while thermal dust emission dominates at higher (millimeter/sub-millimeter) frequencies \citep{planck2016-l01}. Accurately characterizing and removing these foregrounds is therefore critical for unbiased cosmological parameter estimation \citep[see, \eg{}][]{Krachmalnicoff.etal16, Wolz.etal24, Bianchini.etal25}.


While the large-scale foreground emission is relatively well-characterized, small-scale emission remains largely unconstrained by full-sky data. The observational constraints that do exist reveal that the underlying interstellar medium (ISM) and plane-of-sky magnetic field topology are highly complex and structured. For instance, small-scale synchrotron intensity in the Galactic center traces intricate, filamentary magnetic features that are distinctly non-Gaussian \citep{Heywood.etal22, Vanderwoude.etal24}. Similarly, magnetic field lines inferred from dust emission exhibit a similarly non-Gaussian, filamentary morphology \citep{planck2014-XVIII}.

Because these emissions are intrinsically linked to the Galactic magnetic field \citep[\eg{}][]{Rybicki.Lightman79, Lazarian07}, physical 3D simulations are often used to generate mock full-sky observations \citep[\eg{}][]{Waelkens.etal09, Sun.Reich09, Steininger.etal18, Wang.etal20, Porter.etal22}. Even these models remain limited in their ability to resolve small-scale fluctuations, thus, synthetic fields can be employed as a surrogate \citep{Delabrouille.etal13, Thorne.etal17, Vansyngel.etal17, Hervias-Caimapo.Huffenberger22, ThePan-ExperimentGalacticScienceGroup.etal25}. Standard Gaussian approximations overlook the intrinsic non-Gaussianity of the magnetized ISM \citep[\eg{}][]{Biskamp03, Kritsuk.etal07, Kowal.etal07} and current stochastic foreground simulations cannot reproduce these filamentary features. There is therefore no existing method that generates many independent realizations while retaining realistic, non-Gaussian ISM structure.

In this work, we present a framework that fills this gap by generating stochastic, non-Gaussian foreground realizations that reflect the structural complexity expected of Galactic emission \citep{Hervias-Caimapo.Huffenberger22,Abril-Cabezas.etal25,Hervias-Caimapo.etal25,Rahman.etal26}. The framework introduces anisotropic, multifractal, and advected non-Gaussian fluctuations, and has three properties of particular relevance to CMB foreground modelling. First, it is stochastic: it produces arbitrarily many independent realizations, enabling robust characterization of statistical uncertainties. Second, it inherits information from large angular scales, so that the small-scale structure is anchored to the large-scale emission. Third, it is tunable through a small number of physically interpretable parameters, which allows the model to be matched to new high-resolution observations as they become available and to stress-test and identify break points in CMB analysis pipelines for future missions. We use polarized synchrotron emission as an example use case, but the methods apply equally to other foregrounds such as thermal dust emission. This work lays the foundation for new foreground simulation suites with stochastic realizations of realistic, non-Gaussian structure.

Another central challenge is how to quantify the non-Gaussianity \citep[see, \eg{}][]{Burkhart.etal09, Rahman.etal26}. Because the power spectrum (a two-point statistic) is degenerate with respect to various spatial features, it cannot distinguish between a Gaussian random field and a non-Gaussian random field with identical power-law slopes but different phase correlations \citep[\eg{}][]{Chiang.Coles00, Matsubara03, Falceta-Goncalves.etal14}. In this work, our simulated small-scale fields are extrapolated from large-scale WMAP constraints \citep{Bennett.etal03} and assume power-law correlations \citep{Thorne.etal17, ThePan-ExperimentGalacticScienceGroup.etal25}. However, to evaluate the complex phase coherence and morphology introduced by our framework, we must utilize higher-order statistics. Specifically, we evaluate the synthetic fields using Minkowski functionals—which provide an algebraic description of the geometrical and topological properties of a level set \citep{Rahman.etal24, Rahman.etal26}—and structure functions, which probe spatial intermittency and scaling by analyzing the moments of scale-dependent probability density functions \citep{Monin.Iaglom75, Russo.Jury87, Jewell01, Padoan.etal03, Burkhart21}. By analyzing our synthetic fields with these tools, we demonstrate their sensitivity to varying degrees of non-Gaussian structural complexity.

This paper is organized as follows. In \autoref{section:small_scale_models} we describe the generation of small-scale fluctuation fields with anisotropy (\autoref{sec:anisotropies}) and multifractality (\autoref{sec:multifractals}). In \autoref{sec:advection} we introduce a method to generate coherent filamentary structures. Next, in \autoref{section:analysis_methods} we characterize the resulting maps using metrics such as 1D PDFs, Minkowski functionals, structure function moments, and scale filtering. We then extend this framework in \autoref{section:full_sky_models} to construct full-sky synchrotron emission maps by incorprating large-scale emission and modulation fields for Galactic inhomogeneities. Finally, in \autoref{sec:conclusion}, we conclude with a summary.

\section{The Small-Scale Models}
\label{section:small_scale_models}




The \texttt{s6} synchrtron model is availabe in the PySM\footnote{\href{https://github.com/galsci/pysm}{https://github.com/galsci/pysm}} python package. Its synthesis is described in detail in \cite{ThePan-ExperimentGalacticScienceGroup.etal25}, but we will recap some important aspects in this section.

The polarization fraction tensor quantities $i,\,q,\,u$ are used instead of the Stokes $I,\,Q,\,U$ to synthesize non-Gaussian fields.  The transformation from Stokes $I,\,Q,\,U$ to $i,\,q,\,u$ is given by
\begin{subequations}
    \begin{align}
        i &= \frac{1}{2} \ln\bracket{I^2 - P^2},\\
        q &= \frac{1}{2} \frac{Q}{P} \ln\bracket{\frac{I+P}{I-P}},\\
        u &= \frac{1}{2} \frac{U}{P} \ln\bracket{\frac{I+P}{I-P}},
    \end{align}
\end{subequations}%
\noindent where $P = \sqrt{Q^2 + U^2}$ is the usual polarized intensity. \red{We adopt this polarization tensor approach for its compatibility with PySM, which we apply in \autoref{section:full_sky_models}. Nevertheless, the methodology presented here can be readily adapted to raw small-scale $I,\,Q,\,U$ fields with minimal modification.}


The components are generated in the polarization fraction tensor parameterization using
\begin{subequations}\label{eqn:s6_mod}
    \begin{align}
        i &= i_0 + i_\delta i_\mathrm{m},\\
        q &= q_0 + q_\delta q_\mathrm{m},\\
        u &= u_0 + u_\delta u_\mathrm{m},
    \end{align}
\end{subequations}%
\noindent where $i_0,\,q_0,\,u_0$ is the large-scale field on the sky \citep[given by \eg{}the WMAP constraint][]{Bennett.etal03}. The fields $i_\delta,\,q_\delta,\,u_\delta$ are the small-scale realizations where typical methodology generates isotropic, homogeneous Gaussian fluctuations. These small-scale fluctuations are the focus of this paper. The fields $i_\mathrm{m},\,q_\mathrm{m},\,u_\mathrm{m}$ are modulation fields representing inhomogeneous features on the sky. The result of $i_{\delta} i_\mathrm{m},\,q_\delta q_\mathrm{m},\, u_\delta u_\mathrm{m}$ rebalances the power such that small sub-regions match observational constraints and are fields that have larger amplitude fluctuations where the modulation field is strongest, such as in the Galactic plane.

It is upon transformation back to the Stokes' $I,\,Q,\,U$ parameters via
\begin{subequations}
    \label{eqn:backward_pol_tens_transform}
    \begin{align}
        I &= e^i \cosh(\xi),\\
        Q &= \frac{q}{\xi} e^{i} \sinh(\xi),\\
        U &= \frac{u}{\xi} e^{i} \cosh(\xi),
    \end{align}
\end{subequations}%
\noindent where $\xi = \sqrt{q^2 + u^2}$, that small-scale non-Gaussian behavior is obtained.


\subsection{Towards Filamentary Structures -- Anisotropies}
\label{sec:anisotropies}


Our goal is to generate a field capable of stochastically producing structures with a strongly preferred orientation, thereby mimicking the extreme anisotropies revealed by observations of small-scale synchrotron emission \citep{Heywood.etal22}.

We are interested in generating fields $i_\delta,\,q_\delta,\,u_\delta$ that depart from isotropy/homogeneity. To achieve this, we apply the concepts of generalized scale invariance \citep[see, \eg{}][]{Pecknold.etal97, Lewis.etal99, Schertzer.etal13}. Recall that a scale invariant system needs a way of identifying (say, Fourier wave) vectors of the same scale (size) and it needs a way of measuring the scale of these vectors such that no characteristic scale is introduced to the system. Usually, the length of a vector is opted for, and therefore vectors of equal length are of the same scale. As analogues, in spherical harmonics, isotropic scaling only scales $\ell$ and treats all orders $m$ equally. 

Generalized scale invariance moves away from these restrictive cases and states general conditions in which a system can be scale invariant. This is achieved by introducing a spectral-space scale changing operator that warps the harmonic coordinate. Rather than relying on the standard $L_2$ norm, generalized scale invariance defines an anisotropic, direction-dependent ``effective-scale'' metric, which stretches and rotates structures in the harmonic domain.

\begin{figure*}
    \centering
    \includegraphics[scale=1]{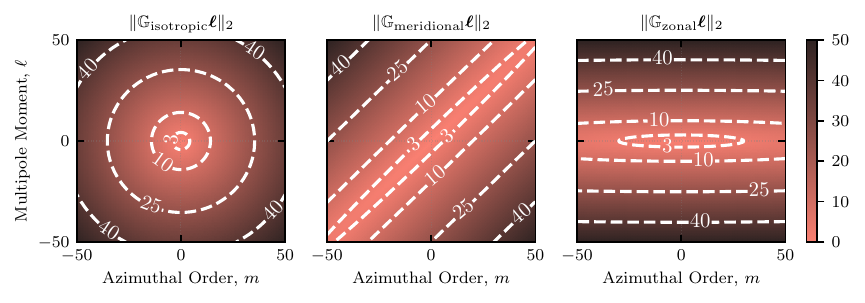}
    \caption{Examples of the effective multipole moments, $\ell_\mathrm{eff} = \lVert \mathbb{G}\vec{\ell} \rVert_2$, extrapolated as a complete $\ell,m$ grid. The rotation and stretching produced by the different transformation matrices are described in \autoref{eqn:matrix_parameters}. The ellipses of the isotropic magnitudes are circular, whereas the meridional and zonal ellipses are stretched and rotated.}%
    \label{fig:grid_transform}%
\end{figure*}

We define a continuous 2D vector $\vec{\ell}$ in the flat-sky harmonic plane, where the vertical axis represents the traditional degree $\ell$ (multipole moment) and the horizontal axis represents the (azimuthal) order $m$:
\begin{align}
    \vec{\ell} = \begin{bmatrix}
        \ell \\
        m
    \end{bmatrix}.
\end{align}
In an isotropic system, the scale or magnitude of this vector is simply the $L_2$ norm:
\begin{align}
    \ell_\mathrm{isotropic} = \lVert \vec{\ell} \rVert_2 = \sqrt{\ell^2 + m^2}.
\end{align}

To introduce directional dependence (anisotropy), we construct a transformation matrix $\mathbb{G}$. This matrix is the product of a scaling tensor $\mathbb{S}$, and a spatial rotation tensor $\mathbb{R}$:
\begin{align}
    \mathbb{G} = \mathbb{S}(s_\ell, s_m)\mathbb{R}(\theta_0),
\end{align}
where
\begin{align}
    \mathbb{S}(s_\ell, s_m) &= \begin{bmatrix}
        s_\ell & 0\\
        0 & s_m \\
    \end{bmatrix},\,
    \mathbb{R}(\theta_0) &= \begin{bmatrix}
        \cos\theta_0 & -\sin\theta_0\\
        \sin\theta_0 & \cos\theta_0
    \end{bmatrix}.
\end{align}
We map the original isotropic harmonic vector $\vec{\ell}$ to a warped coordinate vector $\vec{\ell}^\prime$ via matrix multiplication:
\begin{align}
    \vec{\ell}^{\prime} = \mathbb{G}\vec{\ell}.
\end{align}

Finally, the ``effective scale'' is defined as the magnitude of this transformed vector. This represents the distance from the origin in the warped, anisotropic space:
\begin{align}
    \ell_\mathrm{eff}(\ell,m) = \lVert \vec{\ell}^\prime \rVert_2 = \sqrt{\bracket{\ell^\prime}^2 + \bracket{m^\prime}^2},
\end{align}
which depends on $\ell,m$. By considering $\ell_\mathrm{eff} = \bracket{\mathrm{const.}}$, a tilted and stretched ellipse in the $(\ell,m)$ plane is defined with the ratio of $s_\ell/s_m$ determining the overall level of contraction/stretch in the latitudinal and longitudinal directions, and $\theta_0$ determines the orientation of the anisotropy. Thus, $\mathbb{G} = \mathbb{S}(s_\ell, s_m)\mathbb{R}(\theta_0)$ defines an anistropic filter that alters the spherical harmonic coefficients depending on their orientation. 

Small-scale \emph{anisotropic} and self-similar Gaussian fields are then synthesized from harmonic-space following typical procedures with \eg{}
\begin{align}
    \label{eqn:alm_Cleff}
    a_{\ell,m} = \sqrt{C_{\ell_\mathrm{eff}}^{\mathrm{target}}} \delta^{\mathrm{G}}_{\ell,m},
\end{align}
where $\delta^{\mathrm{G}}_{\ell,m}$ is white variates whose real and imaginary components are independently Gaussian distributed, and $C_{\ell_\mathrm{eff}}^{\mathrm{target}}$ is the power spectrum of the fluctuations interpolated from a target power spectrum $C^{\mathrm{target}}_{\ell}$ onto the effective multipole $\ell_\mathrm{eff}(\ell,m)$ for each $\ell,m$. For exploration purposes, we will parameterize the target power spectrum as
\begin{align}
    \label{eqn:cl_parameterization}
    C_\ell^{\mathrm{target}} = \bracket{\ell^2 + L^2}^{-\alpha/2},
\end{align}
with power-law index $\alpha$, and $L$ a large-scale cutoff/correlation scale.



Note that practically, \autoref{eqn:alm_Cleff} can shift the distribution of power across the $\ell$ modes. Therefore, the estimated isotropic spectrum
\begin{align}
    C_\ell^{\mathrm{isotropic}} = \frac{1}{2\ell + 1} \sum_{m=-\ell}^{\ell} \abs{a_{\ell,m}}^{2},
\end{align}
typically differs from the target power spectrum $C^{\mathrm{target}}_{\ell}$.

To instead generate fluctuations which are anisotropic in real-space, but whose power is distributed equally across $m$ --- so that the isotropic power spectrum equals the target power spectrum --- synthesize the coefficients as follows:
\begin{align}
    a_{\ell,m} = W_{\ell,m} \delta^\mathrm{G}_{\ell,m},
\end{align}
where the total weight $W_{\ell,m}$ describes the power scaling for a specific $(\ell,m)$ mode. This weight is constructed from the product of an anisotropic and an isotropic component:
\begin{align}
    W_{\ell,m} = W_{\ell,m}^{\mathrm{anisotropic}} \times W_{\ell,m}^{\mathrm{isotropic}}.
\end{align}
where the individual scaling factors are defined as
\begin{align}
    W_{\ell,m}^{\mathrm{anisotropic}} &= \sqrt{\frac{C_{\ell_\mathrm{eff}}}{C_{\ell}^{\mathrm{target}}}},\\
    \label{eqn:isotropic_rescale}
    W_{\ell,m}^{\mathrm{isotropic}} &= \sqrt{\frac{C_{\ell}^{\mathrm{target}}}{C_{\ell}^{\mathrm{expected}}}},
\end{align}
and $C_\ell^{\mathrm{target}}$ represents the target power spectrum. The term $C_\ell^{\mathrm{expected}}$ is the expected power introduced by the anisotropic factor, which is obtained by averaging over all $m$ modes:
\begin{align}
    C_\ell^{\mathrm{expected}} = \frac{1}{2\ell+1}\sum_{m=-\ell}^{\ell} C_{\ell_\mathrm{eff}(\ell,m)}.
\end{align}


\begin{figure}
    \centering
    \includegraphics[scale=1]{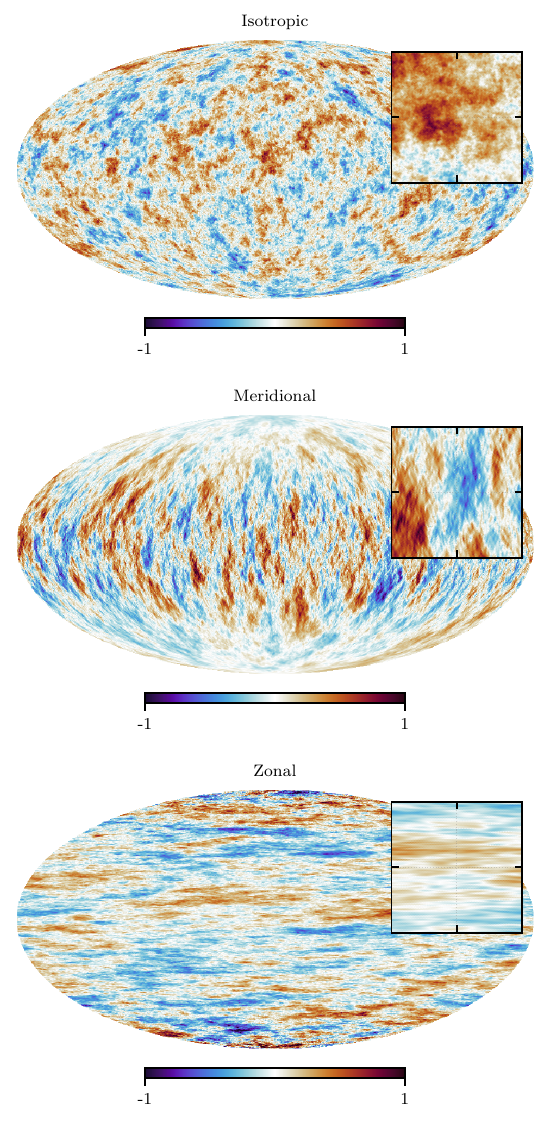}
    \caption{Examples of small-scale temperature fluctuations produced by isotropic (top), meridional (middle) and zonal (bottom) anisotropic transformations parameterized by \autoref{eqn:matrix_parameters}. The fluctuations are generated with the target power spectrum in \autoref{eqn:cl_parameterization} with $L=10$, and $\alpha=8/3$. Inset graphs are a cartesian projection centered at Galactic coordinates $(l,b) = (0^\circ,\,0^\circ)$ with a $14.5^\circ\times14.5^\circ$ field-of-view.}%
    \label{fig:small_scale_examples}%
\end{figure}

For exploration purposes, we describe the following transformation matrices
\begin{subequations}
    \label{eqn:matrix_parameters}
    \begin{align}
        \mathbb{G}_\mathrm{isotropic} &: \curlbracket{s_\ell=1, s_m=1, \theta_0=0^\circ},\\
        \mathbb{G}_\mathrm{meridional} &: \curlbracket{s_\ell=10, s_m=1, \theta_0=45^\circ},\\
        \mathbb{G}_\mathrm{zonal} &: \curlbracket{s_\ell=1, s_m=10, \theta_0=0^\circ},
    \end{align}
\end{subequations}%
\noindent where, as \autoref{fig:grid_transform} shows, $\mathbb{G}_\mathrm{isotropic}$ is an identity matrix and returns $\ell_\mathrm{eff} = \ell_\mathrm{isotropic}$, \ie{}the typical $\ell$'s forming circular magnitudes, $\mathbb{G}_\mathrm{meridional}$ forms $\ell_\mathrm{eff}$ with elliptical magnitudes that are diagonal in the $\ell,m$ plane, and $\mathbb{G}_\mathrm{zonal}$ is elliptical and stretched along the axis of $m$. When considering the true $\ell,m$ space, that for each $\ell > 0$, $m \le \ell$, and for symmetry $m \ge 0$, $\mathbb{G}_\mathrm{meridional}$ favours low-$m$ modes, and $\mathbb{G}_\mathrm{zonal}$ favours high-$m$ modes.

The resultant fluctuations are shown in \autoref{fig:small_scale_examples} with HEALPix\footnote{\href{https://healpix.sourceforge.io}{https://healpix.sourceforge.io}} $N_\mathrm{side}=512$ \citep{Gorski.etal05}, a cutoff scale $L=10$, and $\alpha=8/3$ (\autoref{eqn:cl_parameterization}).\footnote{Note that for the examples in \autoref{section:small_scale_models}, $L$ and $\alpha$ were exaggerated to provide striking visual examples of the methodology. In \autoref{section:full_sky_models}, we consider spectra more relevant to small-scale synchrotron emission.} As the names suggest, the field from $\mathbb{G}_\mathrm{isotropic}$ consists of visually similar fluctuations, $\mathbb{G}_\mathrm{meridional}$ generates visually similar fluctuations elongated North-South, and $\mathbb{G}_\mathrm{zonal}$ generates visually similar fluctuations that are East-West elongated.

An important factor to note is that these modifications generate inhomogeneities in the fluctuations. See, that the ``meridional'' field fluctuations are dominant in the Galactic plane with fluctuations at the (North-South) poles having less amplitude and different scale whereas the ``zonal'' field fluctuations are stronger at the North-South poles.

\subsection{Towards non-Gaussianity -- Multifractality}
\label{sec:multifractals}

Features generated in \autoref{fig:small_scale_examples} are ``visually'' similar. They contain no abrupt fluctuations, singularities, or statistically anomalous regions of high intensity. In other words, we have been generating fields where the variates (\ie{}$\delta^{\mathrm{G}}_{\ell,m}$ in \autoref{eqn:alm_Cleff}) are statistically independent across all $\ell,m$. However, observations indicate the presence of significant levels of non-Gaussianity characterized by structural features whose intensity values lie deep within the non-Gaussian tails of the probability distribution \citep{vonHausegger.etal19, planck2016-l11A, planck2016-l04}.

We now consider a field that is a \emph{cascade} driven by non-linearities that couple different modes and effectively generate a $\delta_{\ell,m}$ that depends non-linearly on both scale and position \citep{Muzy19, Durrive.etal22}. In this scenario, rather than a uniform distribution of fluctuations, the field becomes highly intermittent, characterized by localized bursts, sharp spikes and sudden singularities \citep[see, \eg{}][]{She.Leveque94, Frisch95,Biskamp03}. Given the complex processes that drive and regulate the density and magnetic fields responsible for photon emission, it is entirely justified to stochastically model synchrotron emission as such a cascade \citep{Elmegreen.Scalo04, Lazarian.Pogosyan12}.

We follow \citet{Bacry.etal01} and \citet{Lakhal.etal23} and generate a multifractal field. We describe the process below:
\begin{enumerate}
    \item Generate a correlated noise field, $\omega(\unitvec{n})$, in harmonic space using: $a^{\omega}_{\ell,m} = \sqrt{C_\ell^\mathrm{noise}} \delta^{\mathrm{G}}_{\ell,m}$.\\
    Here, $\delta^{\mathrm{G}}_{\ell,m}$ is complex Gaussian noise and $C_\ell^\mathrm{noise}$ is the power spectrum of $\omega(\unitvec{n})$. Whilst $C_\ell^{\mathrm{noise}}$ does not determine the power spectrum of the resulting multifractal field, it does determine the strength and correlations of the intense regions.
    
    At scales where $\omega$ is uncorrelated, then field will be monofractal. For example, if $C_\ell^{\mathrm{noise}}$ is the same as \autoref{eqn:cl_parameterization}, then the field is monofractal for scales $\ell < L$.
    
    \item Generate log-normal fluctuations by exponentiating the noise field and sampling with noise in physical-space: $h(\unitvec{n}) = e^{\lambda \omega(\unitvec{n})} \delta^{\omega}(\unitvec{n})$.\\
    By exponentiating the field $\omega(\unitvec{n})$, the fluctuations become positive and log-normally distributed. The parameter, $\lambda \in (0,1)$ controls the strength and scaling of the multifractal cascade with large $\lambda$ representing stronger non-Gaussianity and for $\lambda\rightarrow 0$ the field will approach Gaussianity.
    The second noise, $\delta^{\omega}(\unitvec{n})$, is sampled from a Gaussian in real-space for every pixel with variance equal to the variance of $e^{\lambda \omega(\unitvec{n})}$.
    \item Finally, synthesize the multifractal fluctuation field in harmonic-space: $a_{\ell,m} = \sqrt{C_\ell^{\mathrm{target}}} a^{h}_{\ell,m}$.\\
    Here, $a^{h}_{\ell,m}$ is the spherical harmonic coefficients of the field $h(\unitvec{n})$ that was generated in the previous step. The spectrum $C_\ell^{\mathrm{target}}$ describes the scaling of the field.
\end{enumerate}

\begin{figure}
    \centering
    \includegraphics[scale=1]{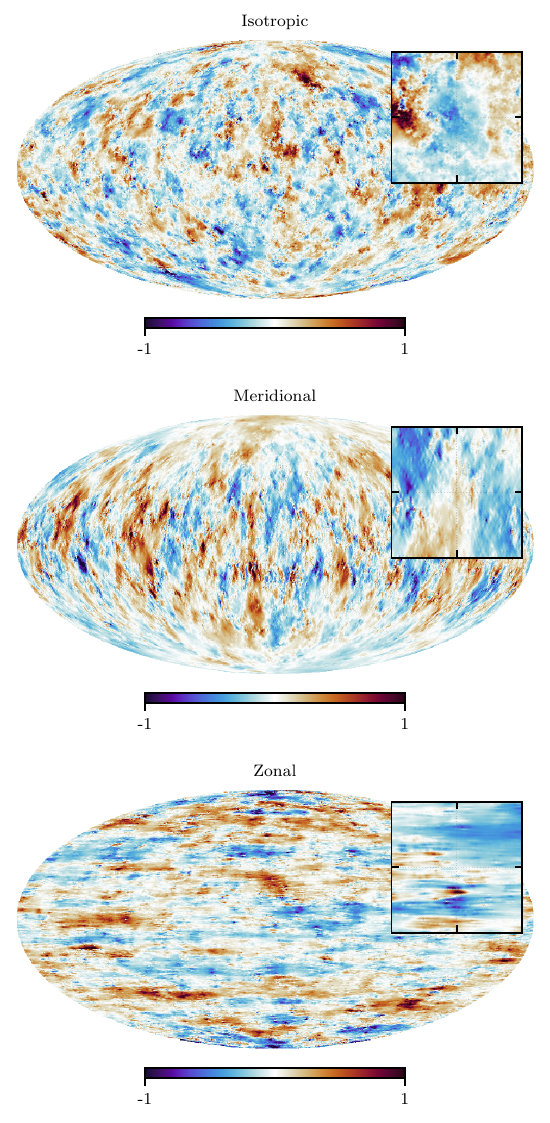}
    \caption{Examples of multifractal ($\lambda = 1/4$) small-scale temperature fluctuations produced by isotropic (top) and meridional (middle) and zonal (bottom) anisotropic transformations parameterized by \autoref{eqn:matrix_parameters}. The fluctuations are generated with the target power spectrum in \autoref{eqn:cl_parameterization} with $L=10$, and $\alpha=8/3$. The overall structure of the anisotropies are the same as \autoref{fig:small_scale_examples}, but the fluctuation amplitudes themselves are \emph{not} ``visually similar'' and there are regions of increased intensity. Inset graphs are a cartesian projection centered at Galactic coordinates $(l,\,b) = (0^\circ,0^\circ)$ with a $14.5^\circ\times14.5^\circ$ field-of-view.}%
    \label{fig:multi_small_scale_examples}%
\end{figure}

We show examples of multifractal fields (with $\lambda = 1/4$) in \autoref{fig:multi_small_scale_examples}. We generate these fields with the same $C_\ell$ parameterization from \autoref{eqn:cl_parameterization} with HEALPix $N_\mathrm{side}=512$ \citep{Gorski.etal05}, $L=10$, and $\alpha=8/3$ for both the correlated noise field, $\omega$, and the final fluctuation field. We also include anisotropies following the procedure we discussed in \autoref{sec:anisotropies}. The visual signature of the multifractal parameter ($\lambda=1/4$) manifests as strong spatial intermittency. Unlike the random fields of \autoref{fig:small_scale_examples} which display a uniform texture, the multifractal fields exhibit heavily localized variance. There are distinct `hotspots' of intense fluctuations interspersed with broad, relatively quiescent `voids'. The distribution and spatial structure of the `hostpots'/`voids' are determined by the correlations of $C_\ell^{\mathrm{noise}}$. Despite the change in distribution, the anisotropies from \autoref{sec:anisotropies} are maintained.


As we will explore in \autoref{sec:structure_functions}, the fields are non-Gaussian. Additionally, the probability density functions depart further from Gaussianity as $\lambda$ increases. Note that the multifractal formalism describes fields whose PDF depends on scale.


\subsection{Building Coherent Structures -- Lagrangian Advection}
\label{sec:advection}

In \autoref{fig:multi_small_scale_examples}, the fluctuations still look like clumpy, disconnected clouds. Even though the power spectrum may be ``correct'', and the higher-order moments successfully generate localized bright spots, is this purely anisotropic multifractal field truly what we would expect the small-scale fluctuations of the interstellar medium to look like? The answer is likely no. The model fails to reproduce thread-like filaments seen in, \eg{}\citet{Heywood.etal22}. The multifractal and anisotropic model lacks this complex spatial coherence. In a purely anisotropic Gaussian model, the phases of the spherical harmonics are random and independent. In the multifractal and anisotropic non-Gaussian model, the phases are indeed coupled --- which is what generates the heavy-tailed intermittency and localized `clumps' --- but they lack the directional coherence needed to form elongated structures. To synthesize filaments, sweeping arcs, and thread-like structures we must generate deeply correlated phases. 

The ISM is a highly turbulent, magnetized plasma. In a highly conductive plasma, the magnetic field lines are ``frozen-in'' to the fluid. As turbulent eddies spin, shear, and cascade energy from large to small scales, they physically stretch and fold the magnetic field lines. It is the filaments that are the physical signatures of magnetic field lines being stretched and sheared by turbulent gas velocities. To emulate this procedure, we adapt the Lagrangian mapping approach \citep{Rosales.Meneveau06, Subedi.etal14, Lubke.etal24}. We generate two fields: an anisotropic multifractal field, and a turbulent velocity field which then advects the multifractal field along the velocity streamlines scale-by-scale. The advection process takes the clumpy, disconnected multifractal ``clouds'' and physically stretches and smears then into filaments. It naturally correlates phases across scales and conserves the heavy-tailed statistics.

\begin{figure}
    \centering
    \includegraphics[scale=1]{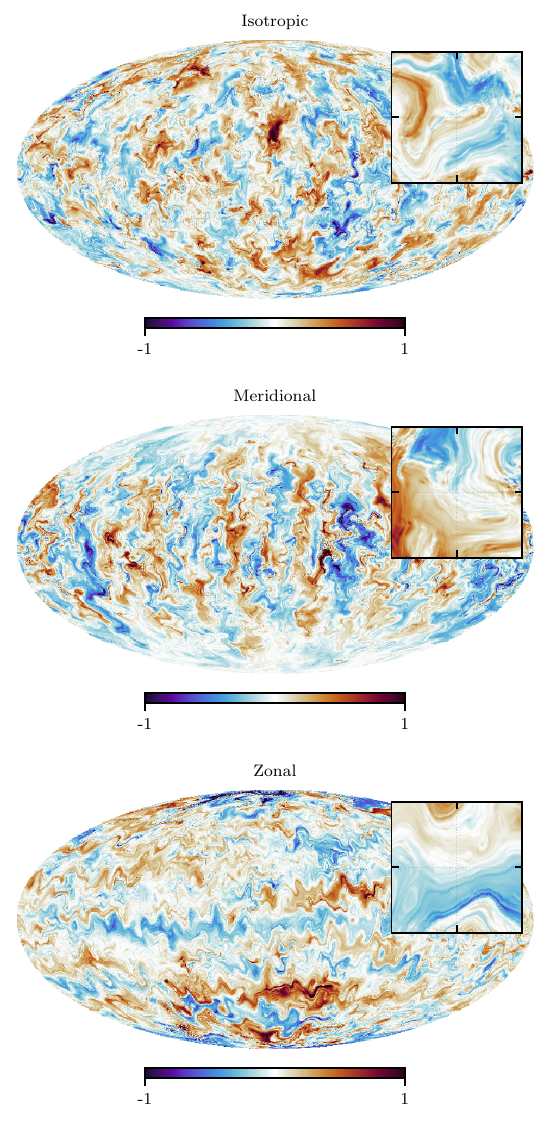}
    \caption{Examples of Lagrangian advected ($c=5$) and multifractal ($\lambda=1/4$) small-scale temperature fluctuations produced by isotropic (top) and meridional (middle) and zonal (bottom) anisotropic transformations parameterized by \autoref{eqn:matrix_parameters}. The fluctuations are generated with the target power spectrum in \autoref{eqn:cl_parameterization} with $L=10$, and $\alpha=8/3$. Inset graphs are a cartesian projection centered at Galactic coordinates $(l,\,b) = (0^\circ,0^\circ)$ with a $14.5^\circ\times14.5^\circ$ field-of-view.}%
    \label{fig:LM_multi_small_scale_examples}%
\end{figure}

We follow the Lagrangian mapping (LM) approach of \citet{Lubke.etal24} and generate a multifractal and geometrically complex random field. We describe the process below:
\begin{enumerate}
    \item Generate a random scalar field: $A(\unitvec{n})$.\\
    Following the anisotropic and multifractal procedures detailed above, we synthesize a random scalar field $A(\unitvec{n})$.
    \item Generate a random second field: $B(\unitvec{n})$.\\
    This second field can also be synthesized using the anisotropic and multifractal procedures.
    \item Derive the turbulent velocity field, $\vec{u}(\unitvec{n})$.\\
    The spatial derivatives of the second field, $B(\unitvec{n})$, take the place of a turbulent velocity field on the sphere:
    \begin{align}
        u_\theta(\unitvec{n}) = \frac{1}{\sin\theta} \frac{\partial B}{\partial \phi}, \quad\quad u_\phi(\unitvec{n}) = -\frac{\partial B}{\partial \theta},
    \end{align}
    where $\vec{u} = (u_\theta, u_\phi)$ with global maximum $u_\mathrm{max} = \mathrm{max}\curlbracket{\sqrt{u_\theta^2 + u_\phi^2}}$.
    \item Generate coordinate displacements: $\Delta \theta(\unitvec{n}),\,\Delta \phi(\unitvec{n})$.\\
    We define the scale-dependent deformation timescale as
    \begin{align}
        \label{eqn:deformation_timescale}
        \tau = \frac{c}{\ell_i u_\mathrm{max}},
    \end{align}
    where $c$ is a free parameter and $\ell_i$ is a representative multipole moment. Using this timescale, the velocity fields perturb the coordinates of $A(\unitvec{n})$ by
    \begin{align}
        \Delta \theta (\unitvec{n}) &= \tau u_\theta(\unitvec{n}),\\
        \Delta \phi (\unitvec{n}) &= \tau \frac{u_\phi(\unitvec{n})}{\sin\bracket{\theta + \Delta\theta}}.
    \end{align}
    It is important to note that this step is scale-dependent, governed by the multipole moment $\ell_i$ in \autoref{eqn:deformation_timescale}. Under the minimal multiscale Lagrangian mapping approach \citep{Subedi.etal14, Lubke.etal24}, large scales successively advect smaller scales. To implement this practically, the initial random scalar fields are decomposed using a sequence of band-pass filters from large to small scales. For each scale iteration $i$, the filter isolates a specific band of spherical harmonics characterized by a representative multipole moment $\ell_i$. This characteristic $\ell_i$ dictates the appropriate deformation timescale for that specific advection step.
    
    In this step, $c$ is a free parameter that must be chosen carefully. Conceptually, $c$ represents the number of eddy turnover times that the ``turbulence'' is allowed to stir/distort the synthetic field at a given scale. In other words, it determines how long the artificial advection is allowed to run before the next scale. Increasing $c$ allows the field to become more twisted and distorted. Initially, this generates prominent coherent structures, though excessive distortion will eventually decorrelate the field entirely.
    \item Advect the random scalar field.\\
    Generate the LM field by advecting the random scalar field, $A(\unitvec{n})$, by sampling/interpolating from itself by the displacements of $\Delta \unitvec{n} = (\Delta \theta(\unitvec{n}),\,\Delta \phi(\unitvec{n}))$:
    \begin{align}
        A(\unitvec{n}) \leftarrow A(\unitvec{n} + \Delta \unitvec{n}).
    \end{align}
    \item Correct the power of the displaced field.\\
    Finally, correct the power of the displacement field in harmonic space:
    \begin{align}
        a^{A}_{\ell,m} \leftarrow W_{\ell,m}^{\mathrm{isotropic}} a^{A}_{\ell,m},
    \end{align}
    where $W_{\ell,m}^{\mathrm{isotropic}}$ is as described in \autoref{eqn:isotropic_rescale}.
\end{enumerate}


Examples of advected fields are shown in \autoref{fig:LM_multi_small_scale_examples}. In these examples, the base field, $A(\unitvec{n})$ is synthesized with: isotropic, meridional, and zonal fluctuations, and the advection field $B(\unitvec{n})$, is synthesized with only isotropic fluctuations. The advection parameters, for all fields are set to $c=5$ and $\lambda=1/4$. Visually, the influence of the Lagrangian advection ($c=5$) is immediately apparent in the form of sheared, filamentary structures that stretch across the fields. Rather than an isotropic distribution of peaks and troughs, the fields exhibit distinct directional streaking and whorl-like patterns indicative of a turbulent flow. The applied anisotropies further elongates these features along a preferred axis.


\subsection{Polarization}
\label{sec:polarization}


\begin{figure}
    \centering
    \includegraphics[scale=1]{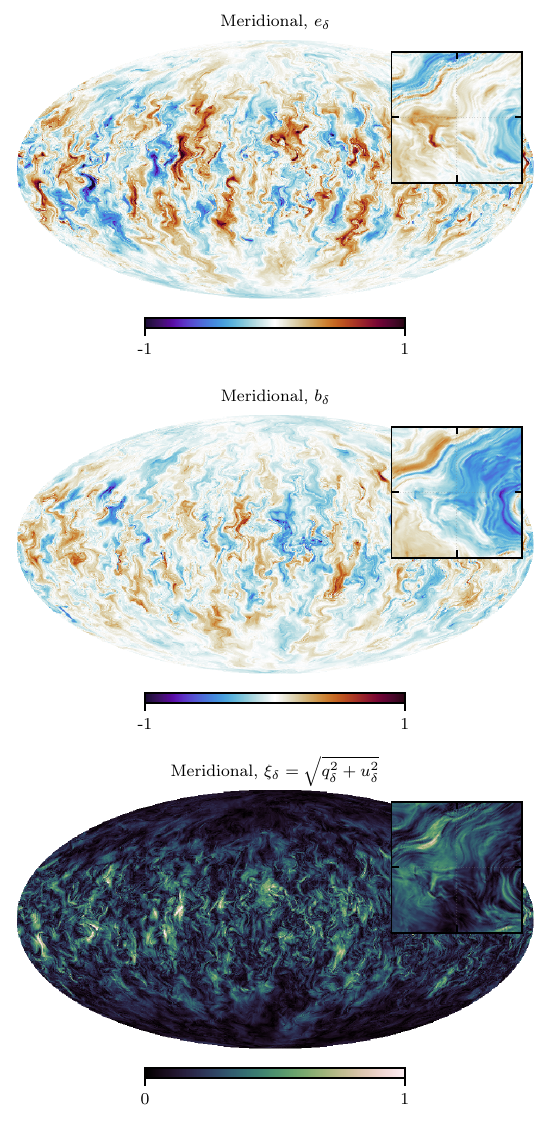}
    \caption{Examples of small-scale $E$-mode (top, $e_\delta$), $B$-mode (middle, $b_\delta$) and polarized intensity (bottom, $\xi_\delta$). The fields are Lagrangian advected ($c=5$) and multifractal ($\lambda=1/4$) with a meridional anisotorpic transformation (\autoref{eqn:matrix_parameters}). The fields are generated with target power spectrum in \autoref{eqn:cl_parameterization} with $L=10$ and $\alpha=8/3$. Inset graphs are a cartesian projection centered at Galactic coordinates $(l,\,b) = (0^\circ,\,0^\circ)$ with a $14.5^\circ\times14.5^\circ$ field-of-view.}%
    \label{fig:pol_small_scale_examples}%
\end{figure}

We generate the linear polarization $q_\delta$, $u_\delta$ using the $e_\delta$, $b_\delta$ pseudo scalar fields (where $e$ and $b$ are the polarization fraction quantities analogous to $E$ and $B$ for $Q$ and $U$). For a covariance matrix described by:
\begin{align}
    \mathbb{A} = \begin{bmatrix}
        C_\ell^{tt} & C_\ell^{te} & C_\ell^{tb}\\
        C_\ell^{te} & C_\ell^{ee} & C_\ell^{eb}\\
        C_\ell^{tb} & C_\ell^{eb} & C_\ell^{tb}\\
    \end{bmatrix}
\end{align}
the Cholesky decomposition ($\mathbb{A} = \mathbb{L} \mathbb{L}^{\intercal}$) is used to generate correlated noise (when non-zero cross-spectra are required). Generally, three independent fluctuation fields will be required, which are then scaled by combinations of the power-spectra described in $\mathbb{A}$ to produce the correct cross-correlations. These three noise fields can be synthesized using the above procedures thus generating polarization fluctuation fields with anisotropies, multifractality, and advected structures.



For a simple example, we provide power spectra as:
\begin{subequations}
    \begin{align}
        C_\ell^{tt} &= C_\ell^{\mathrm{target}}, \quad C_\ell^{ee} = \frac{3}{4} C_\ell^{\mathrm{target}},\quad C_\ell^{bb} = \frac{2}{5} C_\ell^{\mathrm{target}},\\
        C_\ell^{te} &= \frac{3}{10} C_\ell^{\mathrm{target}},\quad C_\ell^{tb} = 0, \quad C_\ell^{eb} = 0,
    \end{align}
\end{subequations}%
\noindent with target $C_\ell^\mathrm{target}$ given by \autoref{eqn:cl_parameterization} with $\alpha=8/3$ and $L=10$. We synthesize polarization fields with multifractal parameter $\lambda=1/4$ and advection strength $c=5$. The resulting polarization fields: $e_\delta$, $b_\delta$, and polarized intensity $\xi_\delta = \sqrt{q_\delta^2 + u_\delta^2}$ are shown in \autoref{fig:pol_small_scale_examples}. Comparing the fields, the higher relative power in the $E$-mode ($C_\ell^{ee} = 3C_\ell^\mathrm{target}/4$) compared to the $B$-mode ($C_\ell^{bb} = 2C_\ell^\mathrm{target}/5$) translates into visibly stronger, higher-constrast structures in $e_\delta$. The polarized intensity map ($\xi_\delta$) reveals a complex structure of highly polarized ridges and depolarization canals where $q_\delta$ and $u_\delta$ approach zero. The polarized intensity becomes small at the poles due to the mathematics of the meridional anisotropic fluctuations. In \autoref{section:full_sky_models} we will explore spatial combinations of maps to correct this artifact. Overall, the visual pattern of anisotropies, non-Gaussianity, and advected structures are displayed in the polarization fields.


\subsubsection{Summary}

\begin{figure*}
    \centering
    \includegraphics[scale=1]{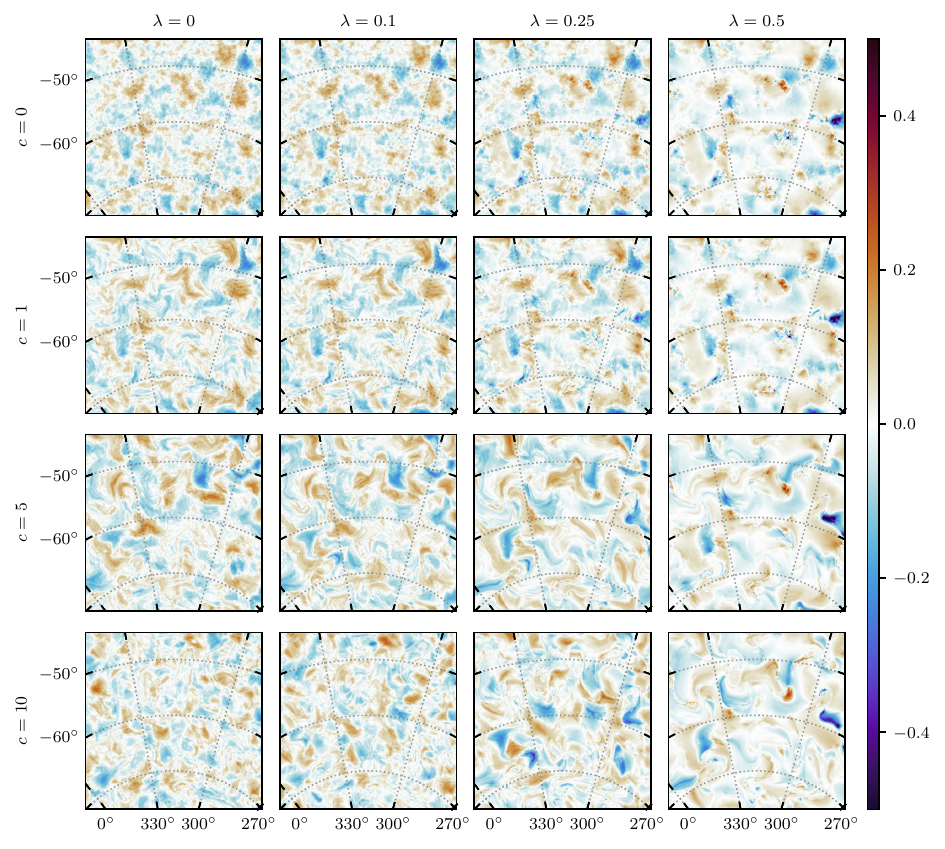}
    \caption{Example sweep of multifractal parameter $\lambda$ (columns, for $\lambda=0,\,0.1,\,0.25,\,0.5$) and advection parameter $c$ (rows, for $c=0,\,1,\,5,\,10$). Gnomonic projection of ($32^\circ \times 32^\circ$) patches of small-scale temperature fluctuations, $i_\delta$, centered at Galactic coordinates $(l,\,b)=(318^\circ,\,-61^\circ)$ with angular resolution of $4.^{\prime}94$. Fluctuations have small-scale power-spectra corresponding to the PySM-\texttt{s6} model.}%
    \label{fig:small_scale_examples_iterate}%
\end{figure*}

In summary, this section has introduced methods to generate complex morphological and statistical fields on the sphere for use in modelling cosmological foreground contamination. Synthesis takes $\sim 2\,\mathrm{min}$ at HEALPix $N_\mathrm{side}=1024$ \citep{Gorski.etal05} on a desktop computer\footnote{All analyses in this paper were performed on a dekstop computer with $32\,\mathrm{GB}$ of RAM, and an $8$ core, $4\,\mathrm{GHz}$ CPU.}. The resulting fields visibly demonstrate the effects of their underlying parameterization. Specifically, the transformation of $(\ell,m)$ by $\mathbb{G}$ (parameterized by $s_\ell,\,s_m,\,\theta_0$) introduces anisotropies that establish a global preference of spatial scale, allowing for the modelling of preferred directional alignments. Additionally, the multifractal nature ($\lambda$) breaks the spatial uniformity of standard Gaussian noise; this produces strong intermittency characterized by clustered high- and low-intensity regions governed by $C_\ell^\mathrm{noise}$. Furthermore, advection from the Lagrangian mapping (LM) generates coherent structures such as filaments, where the advection strength $c$ and the parameters for the flow field determine the spatial displacements. Although this framework introduces several new variables per field, such as $C_\ell^\mathrm{noise}$, $c$, and $\lambda$, alongside the advecting flow field parameters, in practical applications it is reasonable to assume these parameters are identical across the fields to simplify the model.

\autoref{fig:small_scale_examples_iterate} shows additional examples of small-scale temperature fluctuations, $i_\delta$, for varying multifractal and advection strength parameters: $\lambda$ and $c$, respectively. This grid serves as a visual map of this parameter space. Moving across the grid highlights how the spatial clustering driven by $\lambda$ and the coherent structural displacement driven by $c$ couple together to break Gaussian uniformity, providing a robust toolkit for simulating compelx foregrounds.

\textbf{Caveats:} We note the following caveats for those wishing to employ these methods:
\begin{itemize}
    \item The anisotropies (\autoref{sec:anisotropies}) generate stronger amplitude fluctuations in the Galactic plane (for meridional models) and at the Galactic pole (for zonal models). If analyzing sufficiently small sky patches in these regions, the localized power can deviate from that of isotropic fluctuations.
    \item For multifractal fields (\autoref{sec:multifractals}), the nose power-spectrum ($C_\ell^\mathrm{noise}$) controls the distribution and scaling of the intense regions. Therefore, an excessively large effective correlation scale ($L$) in the noise power spectrum ($C_\ell^\mathrm{noise}$) creates large regions of intense fluctuation. Flat-sky analyses confined to these specific regions yield local power spectra that differ from localized patches in other regions of the sky.
    \item Large values of the multifractal parameter ($\lambda \gtrsim 0.75$) inherently introduce singularities that can cause numerical instabilities when transforming between $E$/$B$ and $Q$/$U$ bases. Furthermore, the resulting strong non-Gaussianity can alter the specified variance ($C_\ell^\mathrm{target}$), thereby distorting the intended full-sky power-spectrum.
    \item Setting the advection strength parameter $c$ too high causes the field to become entirely decorrelated.
\end{itemize}


\section{Analysis -- Methods of Measuring Non-Gaussianity}
\label{section:analysis_methods}

All of the fields in \autoref{fig:small_scale_examples}, \autoref{fig:multi_small_scale_examples}, and \autoref{fig:LM_multi_small_scale_examples} have exactly the same power spectrum. This highlights that the power spectrum is degenerate with respect to various spatial features. Given the above developments of anistropies, non-Gaussianity, and coherent structures, we are faced with a challenge: how do we actually measure and characterize their states and significance. In this section, we review and compare several methodologies.



\begin{figure*}
    \centering
    \includegraphics[scale=1]{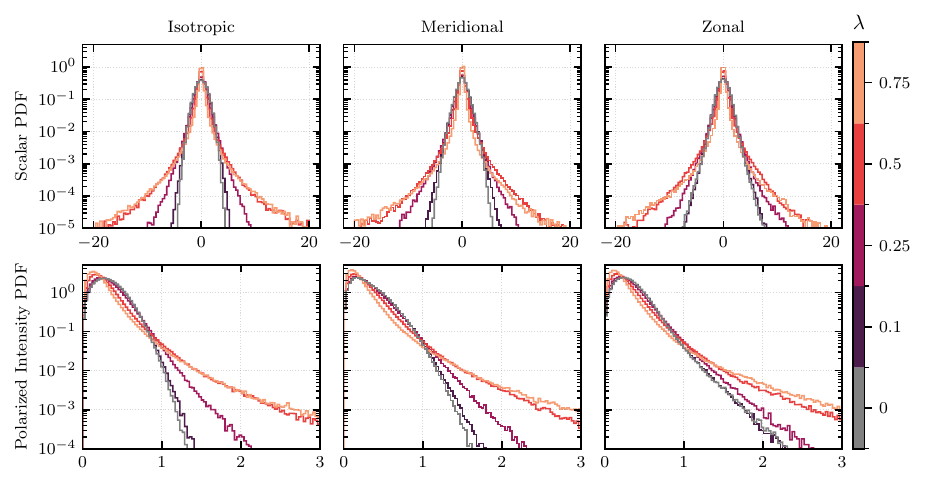}
    \caption{Normalized probability density functions of the standardized non-Gaussian isotropic and anisotropic scalar fields ($i_\delta$; top row) and polarized intensity ($\xi_\delta = \sqrt{q_\delta^2 + u_\delta^2}$; bottom row) with increasing multifractal parameter $\lambda$. Fluctuations are generated with the target power spectrum in \autoref{eqn:cl_parameterization} with $L=25$, $\alpha=7/3$, and no advection ($c=0$).}%
    \label{fig:ani_dists}%
\end{figure*}

\subsection{1D Distributions}


In terms of quanitfying non-Gaussianity, it would seem obvious to first assess whether the probability distribution is indeed Gaussian or non-Gaussian. \autoref{fig:ani_dists} shows the full-sky distributions for isotropic, meridional, and zonal fluctuations for scalar and polarized intensity fluctuations (top and bottom, respectively). It is clear that the distributions become non-Gaussian, and increasingly so for increasing multifractal parameter $\lambda$. For the polarized intensity ($\xi_\delta = \sqrt{u_\delta^2 + q_\delta^2}$) the expected $\chi$ distribution with 2 degrees of freedom becomes increasingly heavy-tailed, departing significantly from this standard Rayleigh profile as the multifractal parameter $\lambda$ increases.


\subsection{Scale-dependent Distributions}
\label{sec:structure_functions}

The multifractal formalism of \autoref{sec:multifractals} implies a scale-dependence to the distribution. To investigate this feature, we use the structure function.

The structure function (SF) is another popular tool for analyzing the statistical properties of fields \citep[see, \eg{}][]{Russo.Jury87, Biskamp03, Cressie93, Frisch95}. The ``angle-averaged $p$-th order structure function'' is defined as the moments of the increments, $\Delta f(\vec{x}, \vec{r}) = f(\vec{x}) - f(\vec{x} + \vec{r})$:
\begin{align}
    \label{eqn:euclidean_SF}
    \overline{S}^{(p)}(r) = \avg{\abs{\Delta f(\vec{x}, \vec{r})}^{p}},
\end{align}
where the angle-brackets denote ensemble averaging as well as performing an average over angles (spherical/circular shells) where $r = \abs{\vec{r}}$. Similar tools have been used in previous analyses for the detection of remaining non-Gaussian signatures in component separated CMB maps \citep[\eg{}][]{Jewell01, Starck.etal04, Miville-Deschenes.etal07}.

In the Gaussian case, we have been using fields where the noise (viz. $\delta^{G}_{\ell,m}$ in \autoref{eqn:alm_Cleff}) has the same distribution for all $\vec{x}$, and the moments of the increments, the structure function (SF), scale as:
\begin{align}
    \overline{S}^{(p)}(r) \propto r^{pH},
\end{align}
where $H$ is the Hurst parameter (a power-law index). Note that the exponent ($pH$) is linear in $p$ and thus $f(\vec{x})$ is \emph{monofractal}. We now consider a field, $g(\vec{x})$, where its increments are instead described by
\begin{align}
    \Delta g(\vec{x}, \vec{r}) = g(\vec{x}) - g(\vec{x} + \vec{r}) = \psi(\vec{r}) \abs{\vec{r}}^{H},
\end{align}
where $\psi(\vec{r})$ is an arbitrary, random, and statistically independent flux-scaling function. This flux-scaling function controls the scaling of $g(\vec{x})$, which is no-longer purely self-similar. The associated SF of $g(\vec{x})$ is
\begin{align}
    \overline{S}^{(p)}(r) = \avg{\abs{\Delta g(\vec{x}, \vec{r})}^{p}} = \avg{\abs{\psi(\vec{r})}^{p}} r^{pH}.
\end{align}
If $\psi(\vec{r})$ has no scale dependence, then $\avg{\abs{\psi(\vec{r})}^{p}} = \bracket{\mathrm{const.}}$ and we retrieve the monofractal behavior described above. However, consider a $\psi(\vec{r})$ scaling such that its statistics follow
\begin{align}
    \avg{\abs{\psi(\vec{r})}^{p}} = \bracket{\frac{L}{r}}^{K(p)},
\end{align}
where $K(p)$ is some moment-dependent function, and $L$ is the largest ``outer-scale'' of the scaling regime. Scaling of $K(p) \ne 0$ implies that the fluctuations of various amplitudes change scale with different exponents (multiscaling) and each realization of such a process is a \emph{multifractal}. Now, for $\overline{S}^{(p)}(r)$ of $g(\vec{x})$:
\begin{align}
    \label{eqn:sf_scaling_parameter}
    \overline{S}^{(p)}(r) \propto r^{\zeta_p},
\end{align}
has scaling $\zeta_p = pH - K(p)$ and is non-linear for convex and non-zero $K(p)$.

The multifractal method of \autoref{sec:multifractals} provides a model for synthesis of a non-Gaussian and multifractal field with log-normal flux model:
\begin{align}
    \ln \psi(\vec{r}) \sim \mathcal{N}(\mu, \sigma^2),
\end{align}
with moments scaling as
\begin{align}
    \avg{\abs{\psi(\vec{r})}^{p}} \propto \bracket{\frac{L}{r}}^{\frac{\lambda}{2}\bracket{p^2 - p}},
\end{align}
and
\begin{align}
    \zeta_p = pH - \frac{\lambda}{2} \bracket{p^2 - p}.
\end{align}
Note that theoretical multifractal scaling is strictly defined over ensemble averages, however, practically the ensemble averages are replaced with spatial averages. Because higher-order moments are dominated by rare, extreme events (intermittency), the spatial average of an individual, finite realization may lack these extremes, resulting in an empirical scaling profile that deviates from ensemble theory \citep{Anselmet.etal84, DudokdeWit04, Zhou12}. Additionally, the LM method (\autoref{sec:advection}) generates coherent structures which alters the probability distribution of extreme increments and fundamentally modifies the multifractal spectrum. This can cause higher-order moments to saturate or grow much slower than the purely multiplicative log-normal model \citep{Rosales.Meneveau08, Subedi.etal14, Lubke.etal24}.


Dissipation structures may not be entirely scale-invariant, particularly, in magnetohydrodynamic turbulence. For example, current sheets can dominate dissipative structures and challenge the scale invariance assumptions \citep{Mininni.Pouquet09, Lesaffre.etal24, Pouquet.etal25}. Nevertheless, the log-normal and advected models are suitable, since, in the first place, there is little observational constraint for $\zeta_p$ on 2D projections of multifractal three-dimensional fluctuations in the ISM. Additionally, the aim is to introduce a model with parameters to control the level of non-Gaussianity for the purposes of testing foreground removal techniques for future CMB missions. 


\begin{figure}
    \centering
    \includegraphics[scale=1]{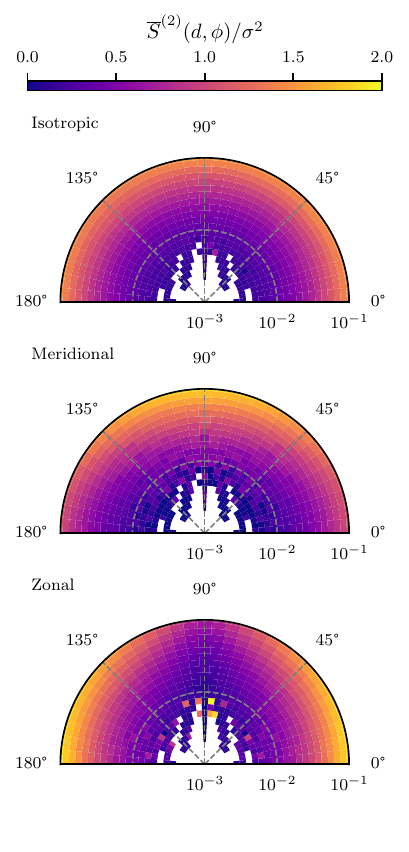}
    \caption{Normalized second order structure functions, $S^{(2)}(d,\phi)/\sigma^2$, for the isotropic (top), meridional (middle), and zonal (bottom) anisotropic fields discussed in \autoref{sec:anisotropies}. These structure functions have dependence on Geodesic distance $d$ and local angle between sampled pairs of points, $\phi$. Fluctuations are generated with target spectrum in \autoref{eqn:cl_parameterization} with $L=10$, $\alpha=8/3$, and no advection ($c=0$).}%
    \label{fig:small_scale_sf2}%
\end{figure}

\subsubsection{Structure Functions}

\begin{figure*}
    \centering
    \includegraphics[scale=1]{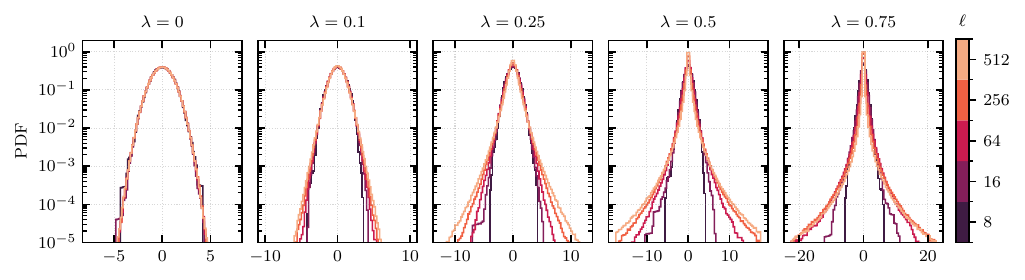}
    \caption{Normalized probability density functions of the standardized band-pass filtered fields at varying filter scales centered at $\ell=8,\,16,\,64,\,256,\,512$. For the monofractal field ($\lambda = 0$), the PDF remains Gaussian across all scales, with only the variance changing. Conversely, for multifractal fields ($\lambda > 0$), the distributions exhibit scale-dependent intermittency, developing increasingly heavy tails and sharper central peaks at smaller scales (larger $\ell$). Fluctuations are generated with the target power spectrum in \autoref{eqn:cl_parameterization} with $L=32$, $\alpha=11/3$, and no advection ($c=0$).}%
    \label{fig:sf_scale_dists}%
\end{figure*}

Since there is an angular power spectrum, and an angular correlation function, it would be natural to extend the relationship to derive some form of angular SF for both scalar and tensor quantities, \eg{}
\begin{align}
    \label{eqn:angular_sf}
    \overline{S}^{(p)}(d, \phi) = \avg{\abs{f(\unitvec{n}_1) - f(\unitvec{n}_2)}^{p}},
\end{align}
where the average is taken over all pairs $(\unitvec{n}_1,\,\unitvec{n}_2)$ separated by Geodesic distance $d$ and oriented at angle $\phi$.


In a flat plane, the increment $\abs{\vec{r}}$ is a straight line. On a sphere, there are two choices for measuring the distance between two locations $\unitvec{n}_1$ and $\unitvec{n}_2$:
\begin{itemize}
    \item Geodesic distance: the actual distance traveling along the curved surface, $d = \arccos\bracket{\unitvec{n}_1 \cdot \unitvec{n}_2}$.
    \item Chord distance (Euclidean distance): The straight-line distance piercing through the sphere, $r = \abs{\unitvec{n}_1 - \unitvec{n}_2} = 2 \sin\bracket{d/2}$.
\end{itemize}
For ``small-scale structure'' (fluctuations of scale $d \ll 1\,\mathrm{radian}$), the sphere locally looks flat, and these two definitions converge ($\sin\bracket{d/2} \approx d/2$, so $r \approx d$).

We estimate\footnote{Due to the computational expense to calculate the entire structure function as a combination of all points with all other points, we sample a fixed number of points from the HEALPix maps for which we evaluate all available pairs. A Python code implementation is available at \href{https://github.com/mab68/small_scale_synch}{https://github.com/mab68/small\_scale\_synch}.} the angular SF of order $p$, parameterized by the Geodesic distance $d$ and the local orientation angle $\phi$ relative to Galactic North. The resulting second-order ($p=2$) angular sSFs for the isotropic and anisotropic fields (introduced in \autoref{sec:anisotropies}) are presented in \autoref{fig:small_scale_sf2}. Each field exhibits a distinct directional signature. As expected, the isotropic field displays no dependence on $\phi$, maintaining a constant amplitude for a given scale $d$. Conversely, the anisotropic fields reveal strong angular dependence. The meridional field, characterized by North-South elongated fluctuations, maintains higher spatial correlations along the longitudinal axis. Because the SF measures the variance of increments, this higher correlation manifests as a suppression of the SF amplitude near $\phi\sim0^\circ$ (and by symmetry $\phi\sim180^\circ$). Similarly, the zonal field is highly correlated East-West, resulting in minimized SF values near $\phi\sim90^\circ$.

\red{Computing the complete anisotropic SFs on the sphere with HEALPix $N_\mathrm{side} \geq 512$ can be computationally expensive \citep{Gorski.etal05}. Thus, an analogous approach is to examine the statistics of the field at a given scale by band-pass filtering it in harmonic space, as is standard in wavelet-based analyses of multifractals \citep[see, \eg{}][]{Muzy.etal91, Muzy.etal93, Arneodo.etal95, Turiel.etal06}. We multiply the spherical harmonic coefficients by an isotropic window $h^{(j)}_\ell$, which selects a narrow, roughly octave-wide range of multipoles around a scale $\ell_j$:
\begin{align}
    \label{eqn:bandpass_filter}
    \Delta f_j(\unitvec{n}) = \sum_{\ell m} h^{(j)}_\ell \, a_{\ell m} \, Y_{\ell,m}(\unitvec{n}).
\end{align}
where $Y_{\ell,m}$ are the spherical harmonics and $a_{\ell m}$ are the spherical harmonic coefficients of the field. The windows are smooth and overlap only with their neighbours, so each $\Delta f_j$ isolates the fluctuations of $f$ at scale $\ell_j$. The filtering is done in harmonic space with \texttt{healpy.almxfl()} \citep{Zonca.etal19}, at the cost of one \texttt{healpy.map2alm()} and one \texttt{healpy.alm2map()} per band. Note that this approach ignores anisotropies (similar to averaging \autoref{eqn:angular_sf} over fixed angles $\phi$), because $h^{(j)}_\ell$ is independent of $m$.}


\autoref{fig:sf_scale_dists} highlights the dependence of the PDF on scale. We band-pass filter the field(s) using \autoref{eqn:bandpass_filter}. For a monofractal (Gaussian, $\lambda=0$) field, the fundamental shape of the distribution is scale-invariant; it remains strictly Gaussian at all scales, entirely characterized by its changing variance. In contrast, the multifractal fields ($\lambda > 0$) exhibit strong scale-dependent intermittency. As the physical scale decreases (larger $\ell$), the distributions progressively depart from a Gaussian profile, developing a sharper central peak and increasingly heavy tails.

\begin{figure}
    \centering
    \includegraphics[scale=1]{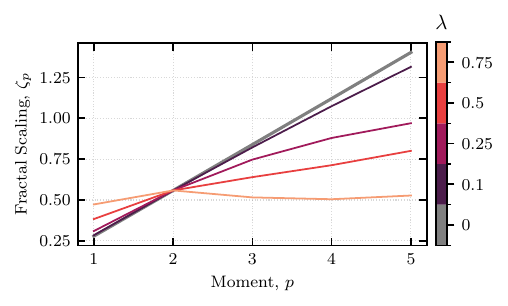}
    \caption{Fractal scaling, $\zeta_p$, estimated using the power-law slopes of the $p$-th order structure function for isotropic fields of increasing multifractal parameter $\lambda$ (and $c=0$, $\alpha=7/3$, $L=25$). The monofractal (Gaussian) field ($\lambda=0$, gray) exhibits linear dependence on $p$ whereas for $\lambda \ne 0$ the behavior is non-linear \ie{}multifractal.}%
    \label{fig:multifractal_scaling}%
\end{figure}

\autoref{fig:multifractal_scaling} shows the scaling parameter $\zeta_p$ (\autoref{eqn:sf_scaling_parameter}) for the isotropic fields across a range of multifractal parameter $\lambda$. For the monofractal Gaussian field ($\lambda=0$), $\zeta_p$ depends linearly on the moment order $p$. In contrast, the multifractal fields ($\lambda>0$) exhibit a distinct non-linear dependence on $p$. The degree of this non-linearity grows with $\lambda$, quantitatively capturing 
the scale-dependent intermittency and increasingly heavy tails --- the transition to non-Gaussianity --- at smaller scales as shown in \autoref{fig:sf_scale_dists}.


The statistics discussed thus far provide a robust characterization of the field's scale dependence and intermittency, but they inherently discard spatial phase information. Consequently, while the isotropic high-pass filtering approach provides a computationally efficient means to evaluate scale-dependent intermittency, it fundamentally integrates out this angular phase information. The 2D angular SF therefore serves as a complementary metric, essential for isolating the global anisotropic signatures that are otherwise lost in 1D scale-filtered statistics. Additionally, the localized anisotropies of coherent structures generated via Lagrangian mapping (\autoref{sec:advection}) tend to average out globally.






\subsection{Minkowski Functionals}

Another method to measure the non-Gaussianity in the maps is the Minkowski functionals \citep[MF;][]{Schmalzing.Gorski98}, which are a common tool to quantify the map-space higher-order statistical properties \citep[\eg{}][]{Chingangbam.Park09, planck2014-a18}. Hadwiger's theorem implies that, for any $D$ dimensional excursion set defined with a threshold value $\nu$:
\begin{align}
    A_\nu = \curlbracket{x \in \mathbb{R}^D \mid f(x) \ge \nu},
\end{align}
there exist $D+1$ MFs that geometrically and topologically describe the morphology of the set \citep{Hadwiger57}. In the case of maps on the sphere $\mathbb{S}^2$, such as temperature or polarized intensity, there are three kinds of MFs: $V_0,\,V_1,\,V_2$, which correspond to the area, the perimeter, and the connectivity of the excursion set, respectively. To compute the MFs we use \texttt{pynkowski}\footnote{\href{https://github.com/javicarron/pynkowski}{https://github.com/javicarron/pynkowski}} \citep{Carones.etal24, CarronDuque.etal24}.

\begin{figure}
    \centering
    \includegraphics[scale=1]{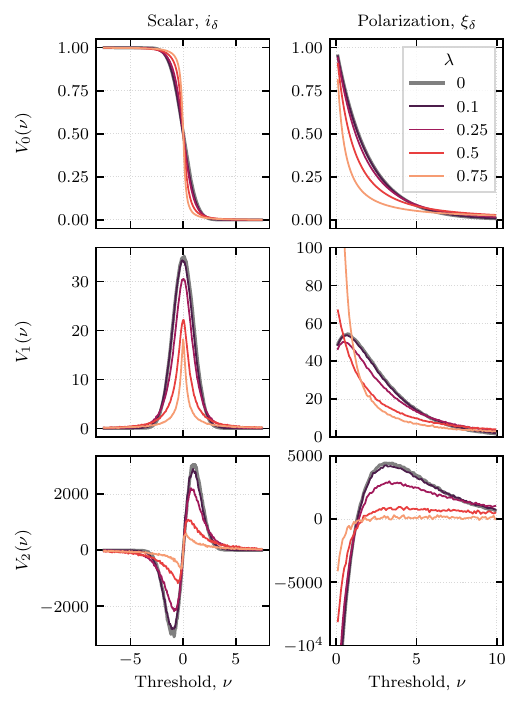}
    \caption{Minkowski functionals $V_0,\,V_1,\,V_2$ for scalar ($i_\delta$, left) and polarized intensity ($\xi_\delta$, right) fluctuations for Gaussian (monofractal, $\lambda=0$; gray) and non-Gaussian (multifractal, with increasing $\lambda$) fields.}%
    \label{fig:iso_minkowski_scalar_P2}%
\end{figure}

We first compute the MFs for the isotropic scalar and polarized intensity fluctuations: $i_\delta,\,\xi_\delta = \sqrt{q_\delta^2 + u_\delta^2}$, respectively. The resulting $V_0,\,V_1,\,V_2$ are shown in \autoref{fig:iso_minkowski_scalar_P2} for Gaussian ($\lambda=0$) and non-Gaussian ($\lambda>0$) fields. The multifractal parameter, $\lambda$, exhibits a clear effect on the MFs for both types of fields (scalar, and polarization). The departure from Gaussianity increases with increasing $\lambda$.

\begin{figure}
    \centering
    \includegraphics[scale=1]{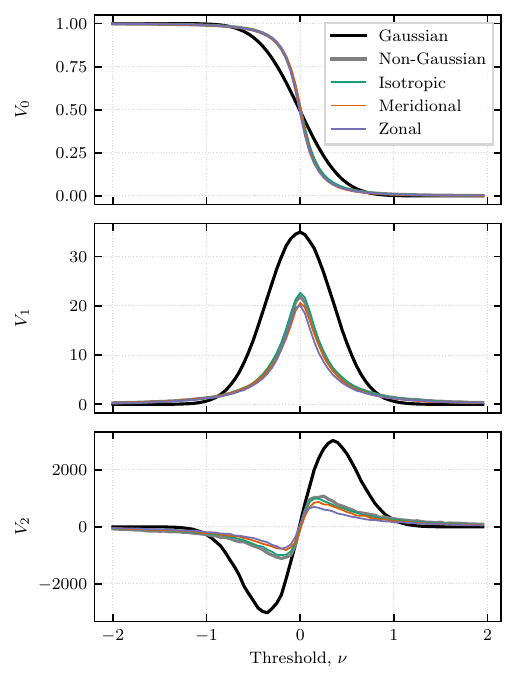}
    \caption{Minkowski functionals $V_0,\,V_1,\,V_2$ for scalar ($i_\delta$, left) fluctuations for fields of different characteristics. An isotropic Gaussian field (monofractal, $\lambda=0,\,c=0$, black), an isotropic non-Gaussian field ($\lambda=1/2,\,c=0$, gray), an isotropic multifractal and advected field ($\lambda=1/2,\,c=5$, green), a meridional multifractal advected field ($\lambda=1/2,\,c=5$, orange), and a zonal multifractal advected field ($\lambda=1/2,\,c=5$, purple).}%
    \label{fig:minkowski_ng_lm_compare}%
\end{figure}

In \autoref{fig:minkowski_ng_lm_compare}, we explore the influence of the full-sky anisotropies on the MFs. There are two clear classes of MF curves: the Gaussian, and the non-Gaussian. The non-Gaussian fields, including those with advected and anisotropic structures, exhibit similar profiles and thus cannot be distinguished. The MFs are not a strong detector for the full-sky anisotropies (\autoref{sec:anisotropies}), nor can they be used to characterize the small-scale coherent structures of advected fields (\autoref{sec:advection}).

Although neither structure functions nor Minkowski functionals are explicitly sensitive to translational phase shifts of LM advection itself, MFs provide a rigorous mathematical framework for extracting the morphology, spatial connectivity, and resulting shape of the intermittent structures.

\subsection{Summary}

Neither structure function or Minkowski function metric can reliably detect properties introduced by the Lagrangian mapping (\autoref{sec:advection}). Further work is required to classify and detect properties of the fields when advection is included (\autoref{sec:advection}). However, examination of multifractal scaling \citep{Lubke.etal24}, or techniques like wavelet scattering transforms \citep{Allys.etal19, Saydjari.etal21, Auclair.etal24}, Minkowski tensors \citep{Schroder-Turk.etal13, Collischon.etal21}, and fourth-order field curvature statistics \citep{Schekochihin.etal02} could hold promise.


It is worth highlighting that the log-normal multifractal model (\autoref{sec:multifractals}) produces a field that becomes \emph{more} non-Gaussian with scale (see, \autoref{sec:structure_functions}). This \red{potentially} contrasts with the results of, \eg{}\cite{Rahman.etal21, Martire.etal23, Rahman.etal24, Rahman.etal26}, in which foreground emission exhibits a kurtosis-dominated non-Gaussian signature that is approximately constant with scale. If this observed trend is a robust feature of the interstellar medium (ISM), it represents a behavior that explicitly cannot be captured by the log-normal or advected multifractal models as currently formulated, in which case, setting $\lambda \approx 0$ would be appropriate. \red{However, because the framework in this paper specifically targets the currently unconstrained small-scale fluctuations, future high-resolution observations will be necessary to definitively test whether this scaling behaviour persists at these scales. Additionally, both Minkowski functionals and structure functions struggle to describe the complete statistical and structural behaviour. This emphasizes the need for further expansion of the available tool box to comprehensively characterize the ISM.}


The framework proposed in this paper provides tools to test methods for quantifying non-Gaussianity and coherent structures. Additionally, when combined with PySM to include data-driven large-scales for full-sky synthesis, it enables robust testing of non-Gaussian foreground pipelines under potential observational conditions.





\section{Synthesizing New Full-sky Simulations}
\label{section:full_sky_models}

\begin{figure*}
    \centering
    \includegraphics[scale=1]{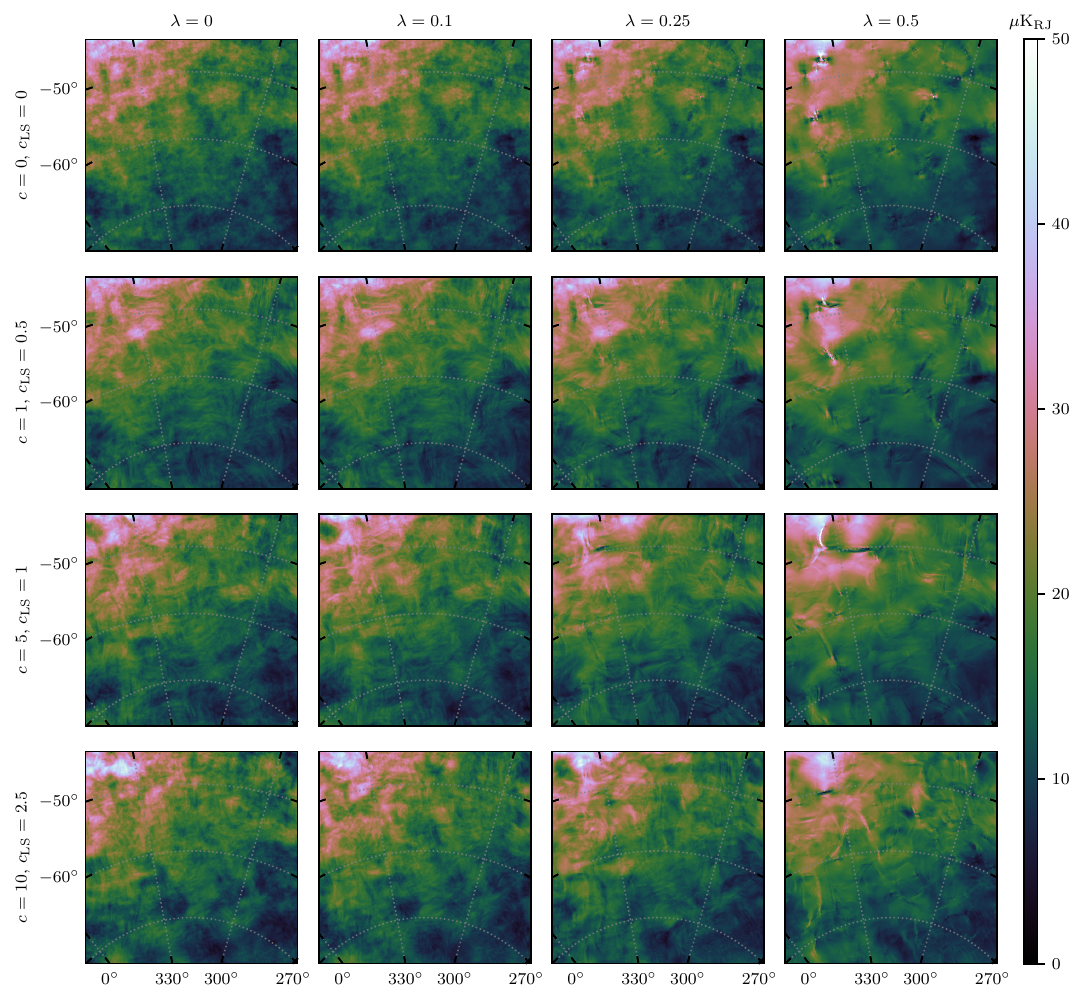}
    \caption{Similar to \autoref{fig:small_scale_examples_iterate}. Example sweep of multifractal parameter $\lambda$ (columns, for $\lambda=0,\,0.1,\,0.25,\,0.5$) and advection parameter $c$ (rows, for $c,c_\mathrm{LS}=(0,0),\,(1,0.5),\,(5,1),\,(10,2.5)$). Gnomonic projection of ($32^\circ \times 32^\circ$) patches of polarized intensity in units of $\mu\mathrm{K}_\mathrm{RJ}$, $P_\mathrm{ref} = \sqrt{Q_\mathrm{ref}^2 + U_\mathrm{ref}^2}$, centered at Galactic coordinates $(l,\,b)=(318^\circ,\,-61^\circ)$ with angular resolution of $4.^{\prime}94$.}%
    \label{fig:examples_iterate_pol}%
\end{figure*}

\begin{figure*}
    \centering
    \includegraphics[scale=1]{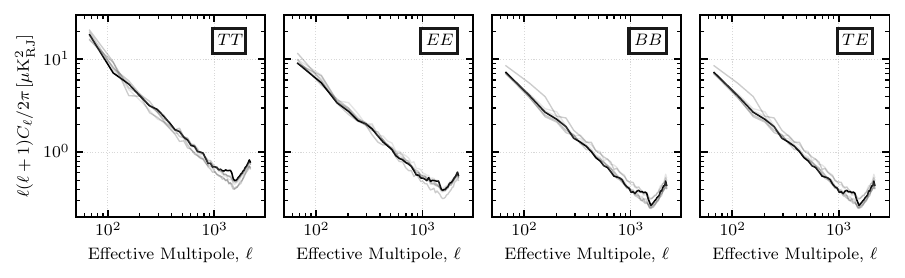}
    \caption{Flat-sky Pseudo-$C_\ell$ computed from the $32^\circ \times 32^\circ$ gnomonically projected patches centered at Galactic coordinates $(l,\,b)=(318^\circ,\,-61^\circ)$ with an angular resolution of $4.^{\prime}94$. These represent the auto- and cross-spectra of the PySM-\texttt{s6} reference fields ($T_\mathrm{ref}$, $E_\mathrm{ref}$, $B_\mathrm{ref}$) across the parameter sweep shown in \autoref{fig:examples_iterate_pol}. The solid black line highlights the default isotropic Gaussian case, while the grey lines indicate the other parameter combinations.}%
    \label{fig:example_highlat_spectra}%
\end{figure*}

In this section, we return to the full-sky (and all-scale) synthesis from \autoref{section:small_scale_models}. To synthesize a mock synchrotron observation, we apply the modulation field and large-scale constraints. Following the synthesis procedures of the PySM-\texttt{s6} model, we use the small-scale power spectra ($C_\ell^\mathrm{target}$), modulation fields ($i_\mathrm{m},\,q_\mathrm{m},\,u_\mathrm{m}$), large-scale fields ($i_0,\,q_0,\,u_0$), and a polarization fraction tensor transform to generate reference $I_\mathrm{ref},\,Q_\mathrm{ref},\,U_\mathrm{ref}$ maps \citep{ThePan-ExperimentGalacticScienceGroup.etal25}. For the \texttt{s6} model, these reference fields correspond to an observation at $23\,\mathrm{GHz}$ that can then be scaled with appropriate power-law index maps to realize fields at other specific observational frequencies.


\begin{figure}
    \centering
    \includegraphics[scale=1]{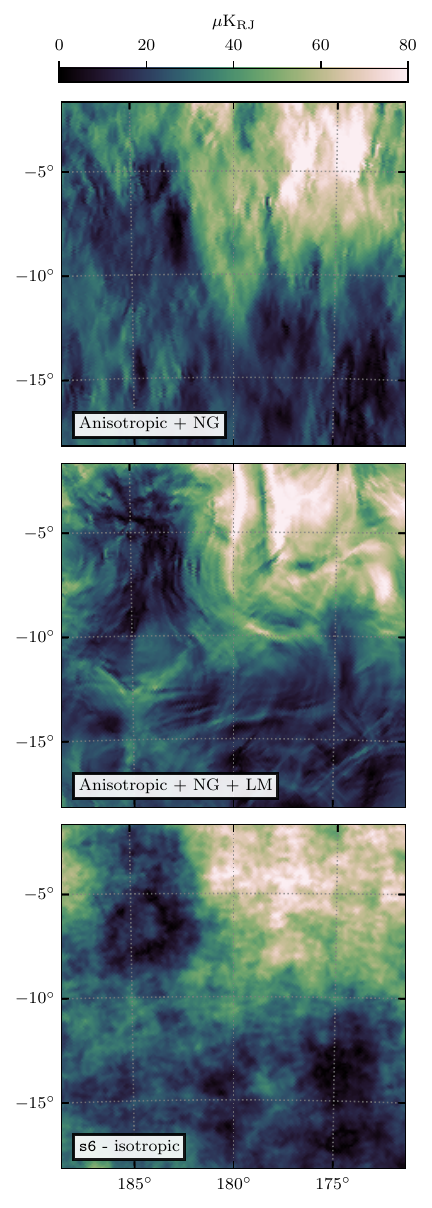}
    \caption{Patches ($16.7^\circ \times 16.7^\circ$) of polarized intensity reference field, $P_\mathrm{ref}$, centered at Galactic coordinates $(l,\,b)=(180^\circ,\,-10^\circ)$ with angular resolution of $4.^{\prime}94$. The reference fields shown correspond to small-scales synthesized with non-Gaussian meridional fluctuations (``Anisotropic + NG'': $\lambda=1/2,\,c=0,\,c_\mathrm{LS}=0$, top), advected meridional non-Gaussian fluctuations (``Anisotropic + NG + LM'': $\lambda=1/4,\,c=5,\,c_\mathrm{LS}=1$, middle), and default PySM-\texttt{s6} isotropic Gaussian fluctuations (``isotropic'': bottom).}%
    \label{fig:P_ref_subfield1}%
\end{figure}

We generate the small-scale temperature and polarization fluctuations using the provided PySM-\texttt{s6} $C_\ell^\mathrm{target}$ spectra, which include non-zero $tt$, $ee$, $bb$, and $te$ correlations. We introduce spatial variation to these fluctuations by applying a spatial filter: the Galactic plane --- defined by the inverse of a $2^\circ$ apodized mask with $f_\mathrm{sky} = 70\%$ from the Planck Legacy Archive\footnote{\texttt{HFI\_Mask\_GalPlane-apo2\_2048\_R2.00.fits}} --- receives meridional anisotropic fluctuations, while the rest of the sky receives isotropic fluctuations. This masking procedure serves a dual purpose: it generates intense, coherent filamentary structures within the Galactic plane, and it rebalances the loss of power at the Galactic poles caused by meridional fluctuations (as seen in \eg{}\autoref{fig:small_scale_examples} and \autoref{fig:multi_small_scale_examples}).

\begin{figure*}
    \centering
    \includegraphics[scale=1]{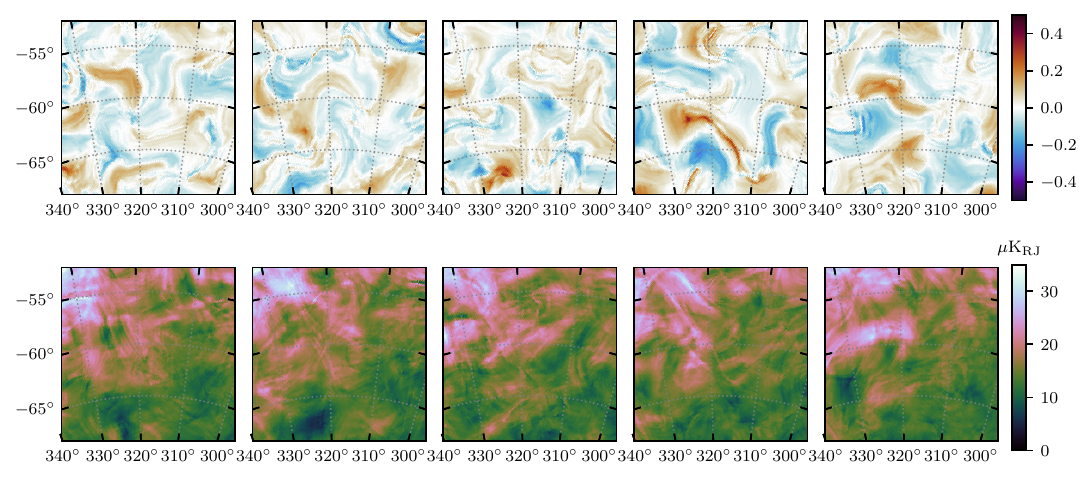}
    \caption{Examples of independently sampled small-scale realizations with $\lambda=1/4,\,c=5,\,c_\mathrm{LS}=2.5$. Gnomonic projection of ($16.7^\circ \times 16.7^\circ$) patches of small-scale temperature fluctuations ($i_\delta$, top) and PySM polarized intensity reference fields ($P_\mathrm{ref} = \sqrt{Q_\mathrm{ref}^2 + U_\mathrm{ref}^2}$, bottom), centered at Galactic coordinates $(l,\,b)=(318^\circ,\,-61^\circ)$ with angular resolution of $4.^{\prime}94$. While each of the five realizations presents a unique spatial morphology, they possess identical statistical properties and characteristic scales across the $i_\delta$ (top) and $P_\mathrm{ref}$ (bottom) fields.}%
    \label{fig:pol_realizations}%
\end{figure*}

We additionally require the small scales to be appropriately coherent with the large scales. By default, there is no phase correlation between the large scales associated with, \eg{}$i_0,\,q_0,\,u_0$ and the small-scale realizations $i_\delta,\,q_\delta,\,u_\delta$. To induce this coherence, we introduce a large-scale deterministic component to the advection. The Lagrangian mapping approach uses a linearized advection equation and accumulates displacements on a scale-to-scale basis. The cascade (small-scale) advection components are calculated as described in \autoref{sec:advection}, and the accumulated displacements which include contributions from a large-scale (LS) field, $A_\mathrm{LS}$, are given by
\begin{align}
    \Delta \theta(\unitvec{n}) &= \bracket{\tau u_\theta(\unitvec{n})} + \bracket{\tau_\mathrm{LS} U_\theta(\unitvec{n})},\\
    \Delta \phi(\unitvec{n}) &= \frac{\bracket{\tau u_\phi(\unitvec{n})} + \bracket{\tau_\mathrm{LS} U_\phi(\unitvec{n})}}{\sin\bracket{\theta + \Delta \theta}}.
\end{align}
The advection components derived from the large scales are:
\begin{align}
    \tau_\mathrm{LS} &= \frac{c_\mathrm{LS}}{\ell U_\mathrm{max}},\\
    U_\theta(\unitvec{n}) &= \frac{1}{\sin \theta} \frac{\partial A_\mathrm{LS}}{\partial \phi},\\
    U_\phi(\unitvec{n}) &= - \frac{\partial A_\mathrm{LS}}{\partial \theta},\\
    U_\mathrm{max} &= \mathrm{max}\curlbracket{\sqrt{U_\theta^2 + U_\phi^2}},
\end{align}
where $c_\mathrm{LS}$ is an additional free parameter representing the large-scale advection strength. For PySM-\texttt{s6} derived fields, the large-scale fields $A_\mathrm{LS}$ correspond to the large-scale constraints divided by the modulation maps: $t_0/t_\mathrm{m},\,e_0/e_\mathrm{m},\,b_0/b_\mathrm{m}$ for the $t_\delta,\,e_\delta,\,b_\delta$ fields, respectively (see, \autoref{section:small_scale_models}). In this manner, the small-scale fields are advected by the large scales and become slightly phase-coherent across the large and small scales, while the large scales remain unmodified. A comparison of the full-sky $Q$ and $U$ maps and their corresponding $C_\ell$ spectra are provided in \autoref{sec:full_sky_comparison}.

\autoref{fig:examples_iterate_pol} shows ``high-latitude'' sub-regions of the polarized intensity reference field, $P_\mathrm{ref} = \sqrt{Q_\mathrm{ref}^2 + U_\mathrm{ref}^2}$, for small-scale models generated with varying multifractal ($\lambda$) and advection ($c,\,c_\mathrm{LS}$) parameters. The top-left panel represents the default PySM-\texttt{s6} isotropic Gaussian fluctuations. Moving across the parameter grid reveals how the spatial clustering driven by $\lambda$ and the coherent structures driven by $c$ couple to break Gaussianity. This demonstrates that the observational imprint on the small scales, as established in \autoref{fig:small_scale_examples_iterate}, is preserved in the full synthesis. 

\autoref{fig:example_highlat_spectra} displays the corresponding pseudo-$C_\ell$ of $T_\mathrm{ref},\,E_\mathrm{ref},\,B_\mathrm{ref}$ for these regions and parameters \citep[estimated using NaMaster;][]{Alonso.etal19}. The spectra of the modified fields (grey lines) closely track the default PySM-\texttt{s6} case ($\lambda=0,\,c=0,\,c_\mathrm{LS}=0$, black line). This reinforces our earlier assertion that two-point correlation functions (power spectra) alone fail to capture the higher-order statistics and phase coherence introduced by the multifractal and advected fields. Furthermore, it confirms that the local power is not significantly altered by the methods introduced in this paper.

\autoref{fig:P_ref_subfield1} focuses on ``low-latitude'' cases of the polarized intensity reference fields. Here, we compare the default \texttt{s6} isotropic Gaussian fluctuations with a model with anisotropic and non-Gaussian fluctuations ($\lambda=1/4,\,c=0,\,c_\mathrm{LS}=0$), as well as one adding advection (``Anisotropic + NG + LM'': $\lambda=1/4,\,c=5,\,c_\mathrm{LS}=1$). Near the Galactic plane, meridional anisotropies impose a strong directional preference on the fluctuations. However, the purely ``Anisotropic + NG'' model retains a somewhat diffuse, cloudy morphology. In contrast, the advected model exhibits filamentary structures, breaking the global homogeneity seen in the purely anisotropic case. At higher latitudes (\eg{}\autoref{fig:examples_iterate_pol}), the spatial masking smoothly replaces anisotropic fluctuations with isotropic ones. Yet, the advected fields continue to display filamentary features with power consistent with those regions of the sky.

A distinct advantage of this methodology is its stochastic nature. In other words, an ensemble of constrained realizations can be generated as opposed to synthesizing small-scale fields from singular observations or simulations. This provides a significant advantage for testing cosmological foreground subtraction methodologies and constraints for future CMB $B$-mode searches \citep[\eg{}][]{Abazajian.etal16, Abazajian.etal19, Hanany.etal19, Abitbol.etal25, Liu.etal25}. \autoref{fig:pol_realizations} shows a sample of five realizations for small-scale $i_\delta$ and PySM-\texttt{s6} polarized intensity reference fields $P_\mathrm{ref}$. As illustrated, the five realizations exhibit visually consistent textures and coherences for both $i_\delta$ and $P_\mathrm{ref}$, confirming they are drawn from the same statistical distribution. However, each individual panel maintains a uniquely randomized spatial configuration. \red{Once again, we emphasize the need to expand the analytical tool box beyond the Minkowski functionals and structure functions in order to adequately characterize such complex, spatially coherent fields.}



In summary, the synthesis framework presented here produces all-scale mock synchrotron skies. By integrating the large-scale observational constraints of PySM with our multifractal and advected small-scale models, we bridge the gap between simple statistical power spectrum matching and generating morphological properties in simulated skies.





\section{Conclusion}
\label{sec:conclusion}


In this paper, we present parameterized mock models of small-scale Galactic intensity and polarization fluctuations, designed specifically to provide rigorous testing grounds for component separation algorithms.


Our framework provides tunable parameters to govern both the morphology and statistical distributions of these fluctuations. Inspired by generalized scale invariance, directional preferences -- or anisotropies -- are introduced via a transformation matrix $\mathbb{G}$ governed by the parameters $s_\ell,\,s_m,\,\theta_0$. Intermittency and non-Gaussianity are controlled using a log-normal multifractal model, which introduces scale-dependent distributions tuned by the multifractal parameter $\lambda$. Finally, phase-coherent structures are generated utilizing a Lagrangian mapping approach, which linearly advects the base field scaled by an advection parameter $c$.

To characterize the impact of these parameters, we evaluated the simulated fields using two complementary staistical metrics: structure functions and Minkowski functionals. While Minkowski functionals successfully capture geometric differences driven by non-Gaussianity in the multifractal fields, they fail to detect directional anisotropies or the small-scale coherent structures introduced by advection. Conversely, structure functions effectively isolate full-sky anisotropies and naturally quantify the multifractal intermittency through the fractal scaling parameter $\zeta_p$. However, when integrated over the full-sky, standard structure functions inherently average out spatial phase information, making them largely blind to localized advected structures.

By integrating our small-scale synthesis pipeline with the large-scale constraints and modulation fields provided by PySM, we succesfully generate realistic, full-sky mock observations exhibiting both non-Gaussian intermittency and coherent filamentary features. The Python implementations for both the fluctuation synthesis and structure function estimators are publicly available at \href{https://github.com/mab68/small_scale_synch}{https://github.com/mab68/small\_scale\_synch}.

Future work will focus on delivering a simulation suite with appropriate realism for CMB foreground subtraction, as well as establishing tighter constraints on the introduced free parameters by comparing our mock maps with high-resolution physical simulations and observational data. Additionally, the introducted methodologies can be similarly applied to, \eg{}thermal dust emission.





\begin{acknowledgments}
Mark A. Bishop is a recipient of the New Zealand Space Scholarship administered by the New Zealand Space Agency on behalf of the New Zealand Government.


\red{Mark A. Bishop utilized Google Gemini 3.1 Pro for minor language editing and Python syntax debugging. The authors reviewed and verified all content and assume full responsibility for the manuscript.}


This research was carried out in part at the Jet Propulsion Laboratory, California Institute of Technology, under a contract with the National Aeronautics and Space Administration (80NM0018D0004).
\end{acknowledgments}

\begin{contribution}
\textbf{MAB:} contributed conceptualization, formal analysis, investigation, methodology, software, validation, visualization, writing -- original draft, writing -- review \& editing.

\noindent\textbf{BSH:} contributed conceptualization, formal analysis, investigation, methodology, validation, visualization, writing -- review \& editing.

\noindent\textbf{KMG:} contributed conceptualization, methodology, writing -- review \& editing.

\noindent\textbf{CL:} contributed conceptualization, methodology, writing -- review \& editing.

\noindent\textbf{GR:} contributed conceptualization, methodology, writing -- review \& editing.
\end{contribution}

\software{Astropy \citep{AstropyCollaboration.etal18},
        HEALPix \citep{Gorski.etal05},
        healpy \citep{Zonca.etal19},
        PySM \citep{Thorne.etal17, Zonca.etal21, ThePan-ExperimentGalacticScienceGroup.etal25},
        NumPy \citep{Harris.etal20},
        Pynkowski \citep{Carones.etal24, CarronDuque.etal24},
        Matplotlib \citep{Hunter07},
        SciPy \citep{Virtanen.etal20},
        Seaborn \citep{Waskom21},
        CMasher \citep{vanderVelden20},
        Kea \citep{Bishop.etal26a},
        NaMaster \citep{Alonso.etal19}.
          }

\appendix
\section{Full-sky Comparison}
\label{sec:full_sky_comparison}


\begin{figure*}
    \centering
    \includegraphics[scale=1]{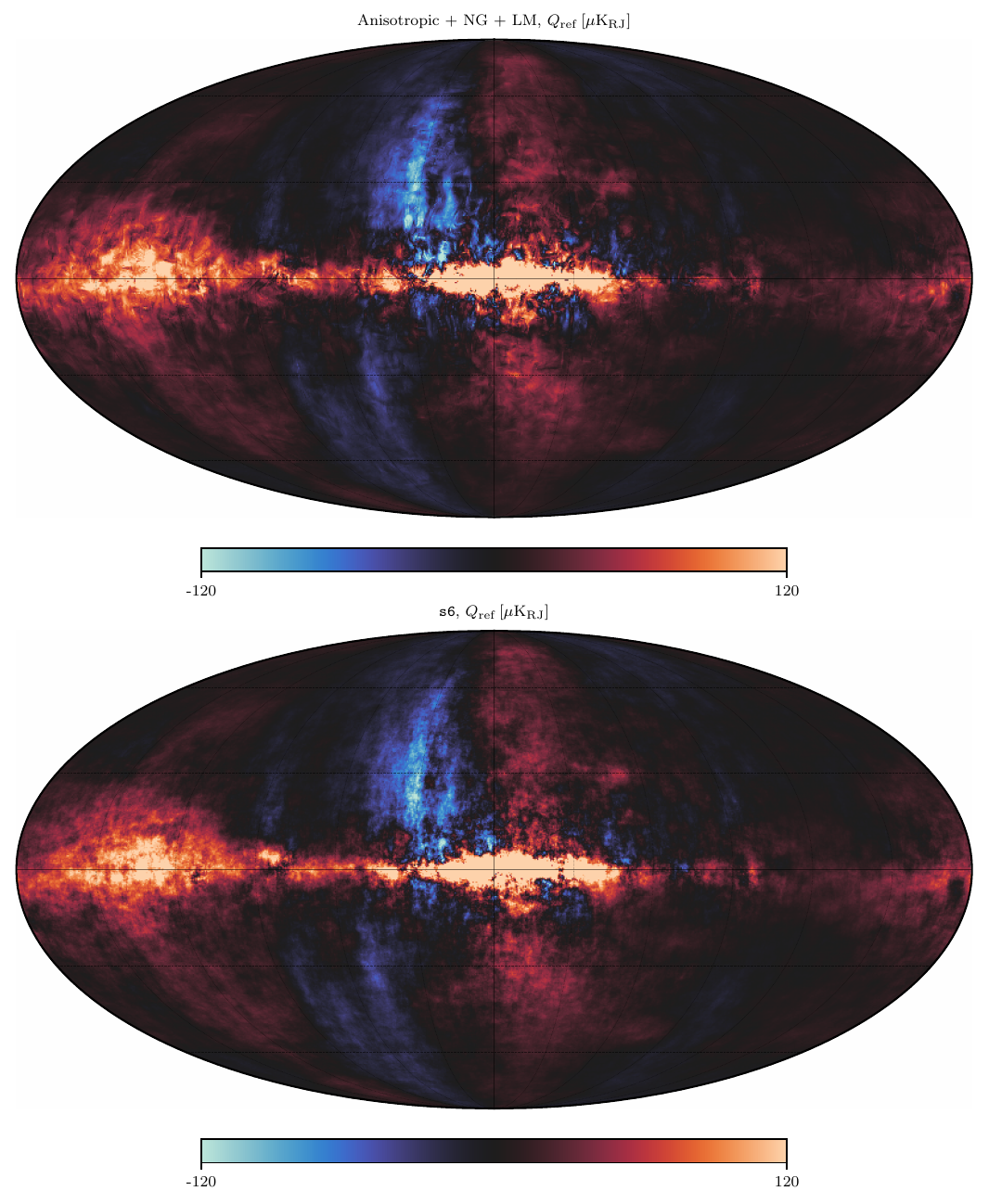}
    \caption{Full-sky PySM reference polarization field $Q_\mathrm{ref}$ in units of $\mu\mathrm{K}_\mathrm{RJ}$ corresponding to a $23\,\mathrm{GHz}$ observation. \textit{Top:} A model with small-scale fluctuations ($i_\delta,\,q_\delta,\,u_\delta$) synthesized with meridional anisotropies in the Galactic plane, multifractality ($\lambda=1/4$), and advection ($c=5,c_\mathrm{LS}=1$). \textit{Bottom:} The default PySM-\texttt{s6} field with small-scale Gaussian and isotropic fluctuations (equivalent to $\lambda=0$, $c=0$).}%
    \label{fig:LM_fullscale_examples_Q}%
\end{figure*}

\begin{figure*}
    \centering
    \includegraphics[scale=1]{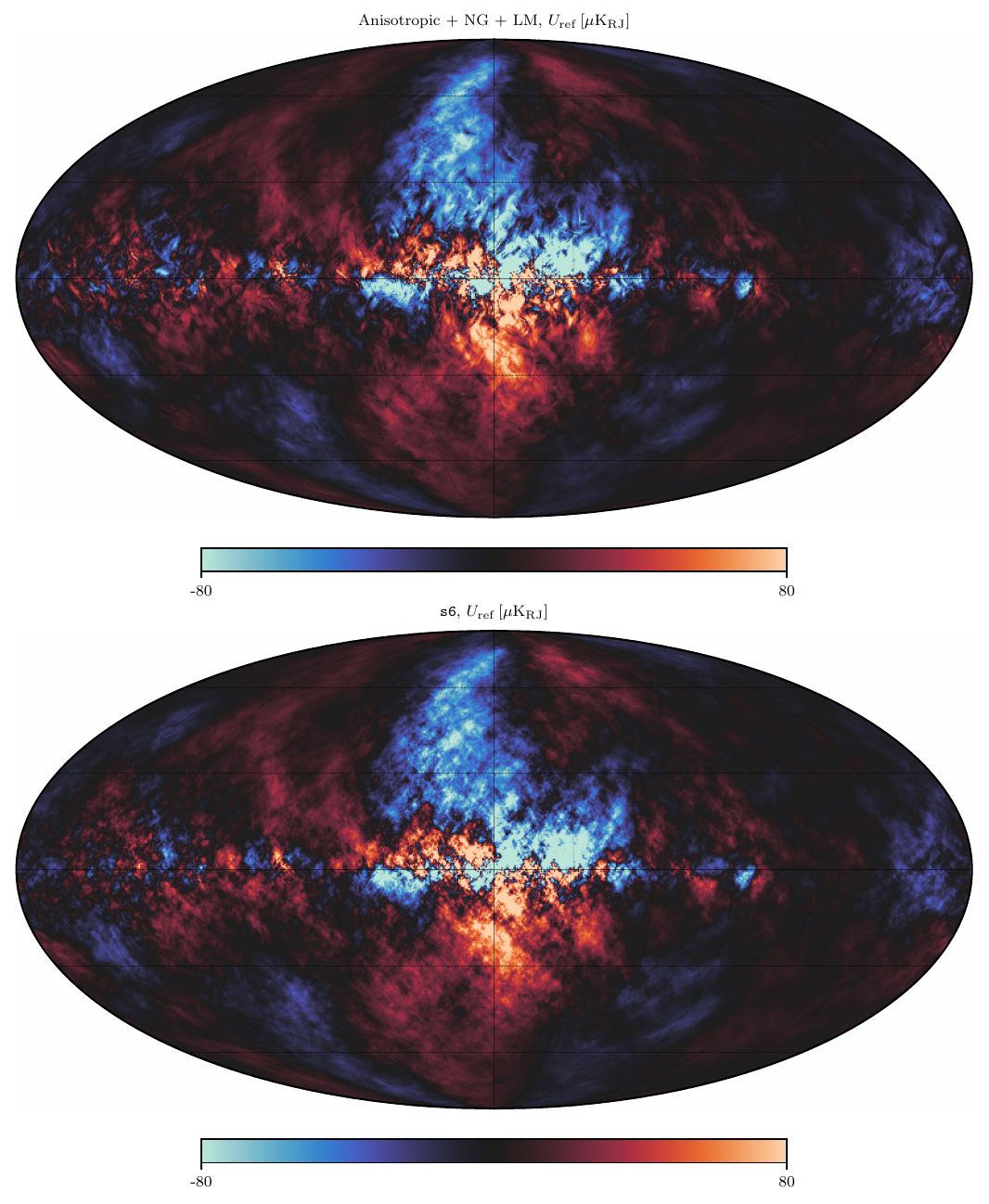}
    \caption{Full-sky PySM reference polarization field $U_\mathrm{ref}$ in units of $\mu\mathrm{K}_\mathrm{RJ}$ corresponding to a $23\,\mathrm{GHz}$ observation. \textit{Top:} A model with small-scale fluctuations ($i_\delta,\,q_\delta,\,u_\delta$) synthesized with meridional anisotropies in the Galactic plane, multifractality ($\lambda=1/4$), and advection ($c=5,c_\mathrm{LS}=1$). \textit{Bottom:} The default PySM-\texttt{s6} field with small-scale Gaussian and isotropic fluctuations (equivalent to $\lambda=0$, $c=0$).}%
    \label{fig:LM_fullscale_examples_U}%
\end{figure*}

Example full-sky fields, synthesized following the procedure discussed in \autoref{section:small_scale_models} and \autoref{section:full_sky_models}, for the $Q$ and $U$ polarization fields are shown in \autoref{fig:LM_fullscale_examples_Q} and \autoref{fig:LM_fullscale_examples_U}, respectively. We compare the base PySM-\texttt{s6} model to a model with anisotropies, multifractality, and coherent advected structures. The large-scale structures in both models exhibit the same patterns and power. However, the small-scale fluctuations of the default \texttt{s6} appear `cloudy' whereas the ``Anisotropic + NG + LM'' model exhibits remarkably more correlated and filamentary structures. Additionally, the anisotropies in the Galactic plane can be seen, with the fluctuation sizes becoming elongated along lines of fixed longitude.

\begin{figure*}
    \centering
    \includegraphics[scale=1]{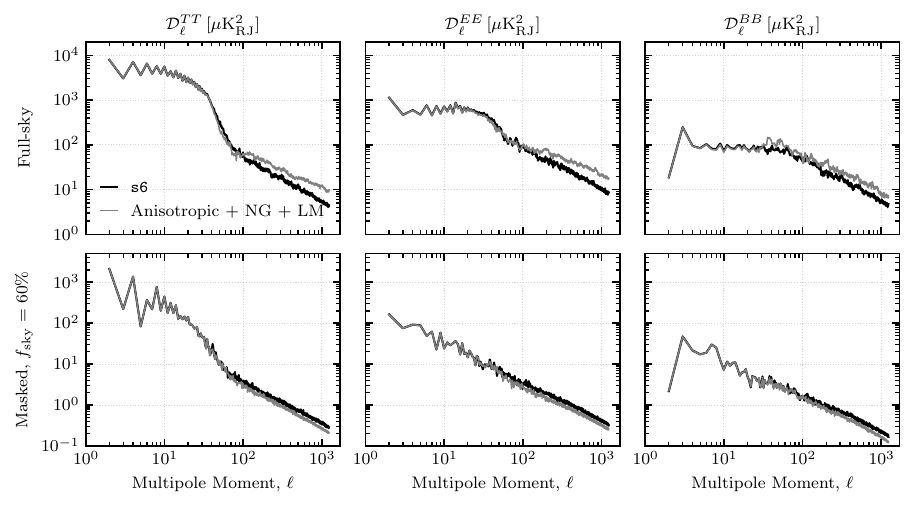}
    \caption{Full-sky auto-power spectra, $\mathcal{D}_\ell = \ell \bracket{\ell+1} C_\ell/ 2\pi$, for the $T_\mathrm{ref},\,E_\mathrm{ref},\,B_\mathrm{ref}$ fields on the full-sky (top row, vis. \autoref{fig:LM_fullscale_examples_Q} and \autoref{fig:LM_fullscale_examples_U}) and masked with $f_\mathrm{sky}=60\%$ (bottom row). Compared is the default PySM-\texttt{s6} model with initial Gaussian isotropic small-scale fluctuations to the ``Anisotropic + NG + LM'' (gray) model with meridional anisotropies in the Galactic plane, multifractality ($\lambda=1/4$), and advection ($c=5$).}%
    \label{fig:fullsky_models_compare_spectra}%
\end{figure*}

The auto-power spectra for the fields are shown in \autoref{fig:fullsky_models_compare_spectra}. Both models have the same large-scale power ($\ell \lesssim 50$). On the small scales ($\ell \gtrsim 50$), the ``Anisotropic + NG + LM'' systematically exhibits slightly more power than the default PySM-\texttt{s6} model by a factor of $\lesssim 3$. This is likely due to the concentration of the fluctuations into the Galactic plane under the meridional anisotropic procedure (\autoref{sec:anisotropies}) and by using the same modulation map as the isotropic model, excess power could be generated at these scales near the Galactic plane. When masking out the Galactic plane using a mask of $f_\mathrm{sky} = 60\%$, the ``Anisotropic + NG + LM'' fluctuations become consistent with the \texttt{s6} model.

\bibliographystyle{aasjournalv7.1}

\bibliography{Planck_bib,References}{}

\end{document}